\documentclass[12pt]{article}
\usepackage{color}
\usepackage{amssymb,amsmath}
\usepackage[dvipdfmx]{graphicx}

\makeatletter
    
    \@addtoreset{equation}{section}
  \makeatother
  
\newcommand{\bea}{\begin{eqnarray}}
\newcommand{\eea}{\end{eqnarray}}
\begin{document}
\setlength{\baselineskip}{0.7cm}
\setlength{\baselineskip}{0.7cm}
\begin{titlepage} 
\begin{flushright}
OCU-PHYS 470\\
YGHP17-08 
\end{flushright}
\vspace*{10mm}
\begin{center}{\Large\bf 
Fermionic Minimal Dark Matter \\
in 5D Gauge-Higgs Unification}
\end{center}
\vspace*{10mm}
\begin{center}
{\large Nobuhito Maru}$^{a}$, 
{\large Nobuchika Okada}$^{b}$, 
and 
{\large Satomi Okada}$^{c}$
\end{center}
\vspace*{0.2cm}
\begin{center}
${}^{a}${\it 
Department of Mathematics and Physics, Osaka City University, \\ 
Osaka 558-8585, Japan}
\\[0.2cm]
${}^{b}${\it Department of Physics and Astronomy, University of Alabama, \\
Tuscaloosa, Alabama 35487, USA} 
\\[0.2cm]
${}^{c}${\it Graduate School of Science and Engineering, Yamagata University, \\
Yamagata 990-8560, Japan} 
\end{center}
\vspace*{1cm}
\begin{abstract} 
\vspace{0.5cm}
We propose a Minimal Dark Matter (MDM) scenario in the context of a simple gauge-Higgs Unification (GHU) model  
   based on the gauge group $SU(3) \times U(1)^\prime$ in 5-dimensional Minkowski space 
   with a compactification of the 5th dimension on $S^1/Z_2$ orbifold. 
A pair of vector-like $SU(3)$ multiplet fermions in a higher-dimensional representation 
   is introduced in the bulk, and the DM particle 
   is identified with the lightest mass eigenstate among the components in the multiplets. 
In the original model description, the DM particle communicates 
   with the Standard Model (SM) particles only through the bulk gauge interaction, 
   and hence our model is the GHU version of the MDM scenario. 
There are two typical realizations of the DM particle in 4-dimensional effective theory: 
(i) the DM particle is mostly composed of the SM $SU(2)_L$ multiplets, or
(ii) the DM is mostly composed of the SM $SU(2)_L$ singlets. 
Since the case (i) is very similar to the original MDM scenario, 
   we focus on the case (ii), which is a realization of the Higgs-portal DM scenario 
   in the context of the GHU model. 
We identify an allowed parameter region to be consistent with the current experimental constraints, 
   which will be fully covered by the direct dark matter detection experiments in the near future. 
In the presence of the bulk multiplet fermions in higher-dimensional $SU(3)$ representations, 
  we reproduce the 125 GeV Higgs boson mass through the renormalization group evolution 
  of Higgs quartic coupling with the compactification scale of $10-100$ TeV.

\end{abstract}
\end{titlepage}

\section{Introduction}

The existence of dark matter (DM) supported by the various cosmological observations
    is  an evidence of physics beyond the Standard Model (SM). 
Clarifying the identities of the DM particle is one of the most important research topics in particle physics and cosmology. 
Among several possibilities, the so-called Weakly Interacting Massive Particle (WIMP) is a prime candidate for the DM. 
Since the WIMP DM is the thermal relic from the early universe, 
   its relic abundance is determined independently of the detailed history of the universe 
   before its decoupling from thermal plasma.  
Currently, many experiments aiming for directly/indirectly detecting DM particles 
  are in operation and planned. 
The discovery of a DM particle may be around the corner.

The GHU scenario \cite{GH} is a unique candidate for new physics beyond the SM, 
   in which the gauge hierarchy problem can be solved without invoking supersymmetry. 
An essential property of the GHU scenario is that the SM Higgs doublet is identified 
   with an extra spatial component of a gauge field in higher dimensions. 
Thanks to the gauge symmetry in higher-dimensions,  
   the GHU scenario, irrespective to the non-renormalizability of the scenario, 
   predicts various finite observables, 
   such as the effective Higgs potential~\cite{1loopmass, 2loop},  
   the effective Higgs coupling with digluon/diphoton \cite{MO, Maru, diphoton}, 
   the anomalous magnetic moment $g-2$~\cite{g-2}, 
   and the electric dipole moment~\cite{EDM}.

Towards the completion of the GHU scenario as new physics beyond the SM, we need to supplement a DM candidate to the scenario. 
In order to keep the original motivation of the GHU scenario to solve the gauge hierarchy problem, 
  the DM candidate to be introduced must be a fermion. 
Since the GHU scenario is defined in a higher dimensional space-time 
  with a gauge group into which the SM gauge group is embedded,  
  it is natural to introduce a DM candidate into the model as a bulk fermion multiplet. 
Hence, the DM particle communicates with the SM particles only through the original bulk gauge interactions. 
This picture is the same as the so-called minimal dark matter (MDM) scenario \cite{MDM}, 
  where the SM $SU(2)_L \times U(1)_Y$ gauge interaction is a unique interaction 
  through which the DM particle communicates with the SM particles.

In our previous work \cite{MMOO}, we have proposed such a MDM scenario in the context of 
  a simple 5-dimensional GHU model based on the gauge group $SU(3)\times U(1)^\prime$ 
  with an orbifold $S^1/Z_2$ compactification of the 5th dimension.   
We have introduced a pair of $SU(3)$ triplet fermions in the bulk 
  along with a bulk mass term and Majorana mass terms on the orbifold fixed point. 
The DM particle is identified as the lightest mass eigenstate, 
  which is a linear combination of $SU(2)_L$ doublets and singlets 
  in the decomposition of the bulk triplets into the SM gauge group of $SU(2)_L \times U(1)_Y$.  
There are two typical cases for a realistic DM scenario: 
  (i) the DM particle is mostly composed of the SM $SU(2)_L$ doublet components, and  
  (ii) the DM particle is mostly composed of the SM $SU(2)_L$ singlets. 
In the case (i),  the DM particle is quite similar to the so-called Higgsino-like neutralino DM in the minimal supersymmetric SM, 
  while the case (ii) is a realization of the so-called Higgs-portal DM scenario in the context of the GHU model. 
We have focused on the case (ii) and identified the allowed parameter region to reproduce the observed DM relic abundance 
  and to satisfy the constraint from the LUX 2016 result \cite{LUX2016} for the direct dark matter search.   
The entire allowed region will be covered by, for example, the LUX-ZEPLIN dark matter experiment \cite{LZ} in the near future. 
We have also shown that in the presence of the bulk $SU(3)$ triplet fermions,  
   the 125 GeV Higgs boson mass is reproduced through the renormalization group (RG) evolution 
   of Higgs quartic coupling with the compactification scale of around $10^8$ GeV. 
However, this compactification scale is too high in the naturalness point of view, and 
   it is desirable to reduce the compactification scale to ${\cal O}$(1 TeV).

Similarly to the MDM scenario \cite{MDM}, we have a variety of choices for representations 
  of the DM multiplets to be introduced. 
In this paper, we extend our previous model by introducing a pair of bulk $SU(3)$ multiplet fermions 
  in higher-dimensional representations, such as ${\bf 6}$-plet, ${\bf 10}$-plet and ${\bf 15}$-plet. 
As has been pointed out in Refs.~\cite{diphoton, MOcolor}, 
  in the presence of such higher representations, 
  the compactification scale can be as low as ${\cal O}$(1 TeV) while reproducing the 125 GeV Higgs boson mass.  
We identify an allowed parameter region to be consistent with the current experimental results.  
For related works on the DM physics in the context of the GHU scenario, see Ref. \cite{GHDM}.

This paper is organized as follows. 
In the next section, we introduce the 5-dimensional GHU model 
   based on the gauge group $SU(3)\times U(1)^\prime$ 
   with an orbifold $S^1/Z_2$ compactification. 
In this context, we propose the GHU version of the MDM scenario,  
   where a DM particle is provided as the lightest mass eigenstate 
   in a pair of $SU(3)$ multiplet fermions introduced in the bulk 
   along with a bulk mass term, brane Majorana mass terms on an orbifold fixed point, 
   and a periodic boundary condition.    
Here, we consider higher-dimensional SU(3) representations, such as 
   ${\bf 6}$-plet, ${\bf 10}$-plet, and  ${\bf 15}$-plet.        
In Sec.~3, we focus on the case that the DM particle communicates with the SM particles  
   through the Higgs boson. 
Calculating the DM relic abundance, we identify an allowed parameter region 
   of the model to reproduce the observed DM density. 
In Sec.~4, we further constrain the allowed parameter region by considering 
   the upper limit of the elastic scattering cross section of the DM particle off nuclei 
   from the current DM direct detection experiments. 
An effective field theoretical approach of the gauge-Higgs Unification scenario 
   will be discussed in Sec.~5, and the 125 GeV Higgs boson mass is reproduced 
   in the presence of the bulk $SU(3)$ triplet fermions with a certain boundary conditions. 
The compactification scale is determined in order to reproduce the Higgs boson mass of 125 GeV. 
The last section is devoted to conclusions.

\section{Fermion MDM in GHU}
We consider a GHU model based on the gauge group 
  $SU(3) \times U(1)^\prime$ \cite{SSS} in a 5-dimensional flat space-time 
  with orbifolding on $S^1/Z_2$ with radius $R$ of $S^1$. 
The boundary conditions should be suitably assigned so as to provide the SM fields as zero modes. 
While a periodic boundary condition corresponding to $S^1$ is applied to all of the bulk SM fields, 
  the $Z_2$ parity for the bulk gauge fields is assigned as
\bea
A_\mu (-y) = P^\dag A_\mu(y) P, \quad A_y(-y) =- P^\dag A_y(y) P,  \quad 
\label{parity}
\eea 
  where $P={\rm diag}(-,-,+)$ is the parity matrix,  and
  the subscripts $\mu$ ($y$) denotes the four (the fifth) 
  dimensional component.  
With this choice of parities, the $SU(3)$ gauge symmetry is 
  explicitly broken down to $SU(2) \times U(1)$. 
A hypercharge is a linear combination of $U(1)$ and $U(1)^\prime$ in this setup. 
For more details of our setup, in particular, bulk fermions including the SM fermions, see Ref.~\cite{MMOO}.

Now we discuss the DM sector in our model. 
In addition to the bulk fermions corresponding to the SM quarks and leptons, 
   we introduce a pair of extra bulk fermions, $\psi$ and $\tilde{\psi}$,  
   which are 6-dimensional, 10-dimensional and 15-dimensional representations 
   under the bulk $SU(3)$ with $U(1)^\prime$ charges of  $2/3$, $1$ and $4/3$, 
   respectively.   
With this choice of the $U(1)^\prime$ charge, 
   the ${\bf 6}$-plet, ${\bf 10}$-plet and ${\bf 15}$-plet bulk fermions include electric-charge neutral components 
   and a linear combination among the charge neutral components serves as the DM particle.    
While we impose the periodic boundary condition in the fifth dimension for all bulk fields, 
   the $Z_2$-parity assignments for the bulk fermions are chosen to be  
\bea
&&\psi(-y) = P \gamma^5 \psi(y), \quad 
\tilde{\psi}(-y) = - P \gamma^5 \tilde{\psi}(y)   \; \; \; ({\rm triplet}), \nonumber \\
&&\psi(-y) = (P \otimes P) \gamma^5 \psi(y), \quad 
\tilde{\psi}(-y) = - (P \otimes P) \gamma^5 \tilde{\psi}(y)  \; \; \; ({\bf 6}{\textrm -}{\rm plet}), \nonumber \\
&&\psi(-y) = (P \otimes P \otimes P) \gamma^5 \psi(y), \quad 
\tilde{\psi}(-y) = - (P \otimes P \otimes P) \gamma^5 \tilde{\psi}(y)  \; \; \; ({\bf 10}{\textrm -}{\rm plet}), \\
&&\psi(-y) = (P \otimes P \otimes P \otimes P) \gamma^5 \psi(y), \quad 
\tilde{\psi}(-y) = - (P\otimes P \otimes P \otimes P) \gamma^5 \tilde{\psi}(y) \; \; \; ({\bf 15}{\textrm -}{\rm plet}) \nonumber . 
\eea 
After the electroweak symmetry breaking, 
  the lightest mass eigenstate among the bulk multiplets is identified with the DM particle.

\subsection{${\bf 6}$-plet MDM}  
Let us first consider the case where the DM provided by a pair of {${\bf 6}$-plet bulk fermions. 
The Lagrangian of the ${\bf 6}$-plets relevant to our DM physics discussion is given by
\bea
{\cal L}_{{\rm DM6}} &=& {\rm Tr} \left[ \overline{\psi}({\bf 6}) \; i D\!\!\!\!/ \; \psi({\bf 6}) 
  + \overline{\tilde{\psi}}({\bf 6}) \;  i D\!\!\!\!/  \; {\tilde \psi}({\bf 6}) )
  - M (\overline{\psi}({\bf 6}) \tilde{\psi}({\bf 6}) 
  + \overline{\tilde{\psi}}({\bf 6}) \psi({\bf 6}) ) \right]
  \nonumber \\
  &&+ \delta(y) \left[ 
  \frac{m}{2} \overline{\nu_{sR}^{(0)c}} \nu_{sR}^{(0)} 
  +\frac{\tilde{m}}{2} \overline{\tilde{\nu}_{sL}^{(0)c}} \tilde{\nu}_{sL}^{(0)} 
  + {\rm h.c.} \right], 
\label{DMLagrangian6}
\eea
 where $D\!\!\!\!/ $ is the covariant derivative for the ${\bf 6}$-plets. 
The bulk $SU(3)$ ${\bf 6}$-plet is expressed in a matrix form
\bea
\psi({\bf 6})=
\left(
\begin{array}{ccc}
\Sigma_\uparrow & \frac{1}{\sqrt{2}} \Sigma_0 & \frac{1}{\sqrt{2}} \nu \\
\frac{1}{\sqrt{2}} \Sigma_0 & \Sigma_\downarrow & \frac{1}{\sqrt{2}} e \\
\frac{1}{\sqrt{2}} \nu & \frac{1}{\sqrt{2}}e & \nu_s \\
\end{array}
\right), 
\label{rep6-1}
\eea
where the quantum numbers for $SU(2)$ representation, 
 the third component of the isospin $I_3$, and the electric charge $Q_{{\rm em}}$ 
 of various fields in the matrix are 
\bea
\begin{array}{lll}
\Sigma_\uparrow({\bf 3}, 1, 2), & 
\Sigma_0({\bf 3}, 0, 1), & 
\Sigma_\downarrow({\bf 3}, -1, 0), \\
\nu({\bf 2}, 1/2, 1), & 
e({\bf 2}, -1/2, 0), &
\nu_s({\bf 1}, 0, 0). \\
\end{array}
\label{rep6-2}
\eea
The corresponding mirror fermion ${\bf 6}$-plet $\tilde{\psi}({\bf 6})$ takes a similar form 
 with tilde for all fields.

With the non-trivial orbifold boundary conditions, 
   the ${\bf 6}$-plet fermions are decomposed into 
   the SM $SU(2)$ triplet, doublet and singlet fermions. 
As we will see later, the DM particle is provided as a linear combination of 
   the $SU(2)$ singlets and the electric charge neutral components in the $SU(2)$ doublets and triplets. 
In Eq.~(\ref{DMLagrangian6}) we have introduced a bulk mass ($M$) to avoid exotic massless fermions.  
Here, we have also introduced Majorana mass terms on the brane at $y=0$ 
   for the zero-modes of the components of the ${\bf 6}$-plets ($(\nu_s^{(0)})_R$ and $(\tilde{\nu}_s^{(0)})_L$),   
   which are singlet under the SM gauge group. 
The superscript ``$c$" denotes the charge conjugation. 
With the Majorana masses on the brane, the DM particle in 4-dimensional effective theory 
   is a Majorana fermion, and hence its spin-independent cross section with nucleons 
   through the $Z$-boson exchange vanishes in the non-relativistic limit.

Let us focus on the following terms in Eq.~(\ref{DMLagrangian6}), 
   which are relevant to the mass terms in 4-dimensional effective theory: 
\bea
{\cal L}_{{\rm mass6}} &=& 
 {\rm Tr} \left[ \overline{\psi} i \Gamma^5 (\partial_y - 2 ig \langle A_y \rangle) \psi 
 + \overline{\tilde{\psi}} i \Gamma^5 (\partial_y -2 ig \langle A_y \rangle) \tilde{\psi} 
 - M(\overline{\psi} \tilde{\psi} + \overline{\tilde{\psi}} \psi) \right]
 \nonumber \\
 &&+ \delta(y) \left[ \frac{m}{2} \overline{\nu_{sR}^{(0)c}} \nu_{sR}^{(0)} 
 +\frac{\tilde{m}}{2} \overline{\tilde{\nu}_{sL}^{(0)c}} \tilde{\nu}_{sL}^{(0)}  
 + {\rm h.c.} \right], 
\label{L6_mass}
\eea
 where $\Gamma^5= i \gamma^5$. 
Expanding the bulk fermions in terms of KK modes as
\bea
\psi(x,y) &=& \frac{1}{\sqrt{2\pi R}} \psi^{(0)}(x) 
+ \frac{1}{\sqrt{\pi R}} \sum_{n=1}^\infty \psi^{(n)}(x) \cos \left(\frac{n}{R} y\right) 
\nonumber \\
&& ({\rm for}~\Sigma_{\uparrow R, 0R, \downarrow R}, \nu_L, e_L, \nu_{sR}, 
 \tilde{\Sigma}_{\uparrow L, 0L, \downarrow L}, \tilde{\nu}_R, \tilde{e}_R, \tilde{\nu}_{sL}), \\
\psi(x,y) &=& \frac{1}{\sqrt{\pi R}} \sum_{n=1}^\infty \psi^{(n)}(x) \sin \left(\frac{n}{R} y\right)
\nonumber \\
&& ({\rm for}~\Sigma_{\uparrow L, 0L, \downarrow L}, \nu_R, e_R, \nu_{sL}, 
 \tilde{\Sigma}_{\uparrow R, 0R, \downarrow R}, \nu_L, e_L, \nu_{sR}), 
\eea
and integrating out the fifth coordinate $y$, we obtain the expression in 4-dimensional effective theory.  
The zero-mode parts for the electric-charge neutral fermions are found to be 
\bea
{\cal L}^{\rm zero-mode}_{\rm mass} 
&=& 
 i \frac{m_W}{\sqrt{2}} \left[ 
 \overline{e_L^{(0)}} (\Sigma_{\downarrow R}^{(0)} + \nu_{sR}^{(0)}) 
 + (\overline{\tilde{\Sigma}_{\downarrow L}^{(0)}} + \overline{\tilde{\nu}_{sL}^{(0)}}) \tilde{e}_R^{(0)} 
 + {\rm h.c.} \right] 
 \nonumber \\
 && -M \left( \overline{\Sigma_{\downarrow R}^{(0)}} \tilde{\Sigma}_{\downarrow L}^{(0)} 
 + \overline{\nu_{sR}^{(0)}} \tilde{\nu}_{sL}^{(0)} + \overline{\tilde{e}_R^{(0)}} e_L^{(0)} 
+{\rm h.c.} \right)  
\nonumber \\
&& +
\left( \frac{m}{2} \overline{\nu_{sR}^{(0)c}} \nu_{sR}^{(0)} 
 + \frac{\tilde{m}}{2} \overline{\tilde{\nu}_{sL}^{(0)c}} \tilde{\nu}_{sL}^{(0)} 
 + {\rm h.c.} \right) 
\nonumber \\
& \to & 
 - \frac{m_W}{\sqrt{2}}  \left[ 
 \overline{e_L^{(0)}} (\Sigma_{\downarrow R}^{(0)} + \nu_{sR}^{(0)}) 
 + (\overline{\tilde{\Sigma}_{\downarrow L}^{(0)}} + \overline{\tilde{\nu}_{sL}^{(0)}}) \tilde{e}_R^{(0)} 
 + {\rm h.c.} \right] 
 \nonumber \\
 && -M \left( \overline{\Sigma_{\downarrow R}^{(0)}} \tilde{\Sigma}_{\downarrow L}^{(0)} 
 + \overline{\nu_{sR}^{(0)}} \tilde{\nu}_{sL}^{(0)} + \overline{\tilde{e}_R^{(0)}} e_L^{(0)} 
+{\rm h.c.} \right)  
\nonumber \\ 
&& -
\left( \frac{m}{2} \overline{\nu_{sR}^{(0)c}} \nu_{sR}^{(0)} 
 + \frac{\tilde{m}}{2} \overline{\tilde{\nu}_{sL}^{(0)c}} \tilde{\nu}_{sL}^{(0)} 
 + {\rm h.c.}  \right), 
\eea
where $m_W=gv/2$ is the $W$-boson mass, and the arrow means the phase rotations 
\bea
\nu_s^{(0)} \to i \nu_s^{(0)}, \quad 
\tilde{\nu}_s^{(0)} \to i \tilde{\nu}_s^{(0)}, \quad  
\Sigma_\downarrow^{(0)} \to i \Sigma_\downarrow^{(0)}, \quad 
\tilde{\Sigma}_\downarrow^{(0)} \to i \tilde{\Sigma}_\downarrow^{(0)}.  
\label{phaserot}
\eea
It is useful to rewrite these mass terms in a Majorana basis defined as  
\bea
&&\chi \equiv \nu^{(0)}_{sR} + \nu^{(0)c}_{sR}, \quad 
\tilde{\chi} \equiv \tilde{\nu}^{(0)}_{sL} + \tilde{\nu}^{(0)c}_{sL}, 
\nonumber \\
&&\omega \equiv e^{(0)}_{L} + e^{(0)c}_{L}, \quad 
\tilde{\omega} \equiv \tilde{e}^{(0)}_{R} + \tilde{e}^{(0)c}_{R}, 
\nonumber \\
&& \eta \equiv \Sigma_{\downarrow R}^{(0)} + \Sigma_{\downarrow R}^{(0)c}, \quad
\tilde{\eta} \equiv \tilde{\Sigma}_{\downarrow L}^{(0)} + \tilde{\Sigma}_{\downarrow L}^{(0)c}, 
\label{Majorana}
\eea
and we then express the mass matrix (${\cal M}_{{\rm N6}}$) as 
\bea
{\cal L}_{6{\rm mass}}^{\rm zero-mode} 
 & &= 
-\frac{1}{2}(
\begin{array}{cccccc}
\overline{\chi} & \overline{\tilde{\chi}} & \overline{\omega} & \overline{\tilde{\omega}} & \overline{\eta} & \overline{\tilde{\eta}} \\
\end{array}
)
{\cal M}_{{\rm N6}}
\left(
\begin{array}{c}
\chi \\
\tilde{\chi} \\
\omega  \\
\tilde{\omega }\\
\eta \\
\tilde{\eta} \\
\end{array}
\right)  \nonumber \\
& &=
 -\frac{1}{2}(
\begin{array}{cccccc}
\overline{\chi} & \overline{\tilde{\chi}} & \overline{\omega} & \overline{\tilde{\omega}} & \overline{\eta} & \overline{\tilde{\eta}} \\
\end{array}
) \times \nonumber \\
&& 
\left( 
\begin{array}{cccccc}
m      & M            &  \sqrt{2} m_W  & 0 & 0 & 0 \\
M      & {\tilde m} &   0       & - \sqrt{2} m_W & 0 & 0 \\
\sqrt{2} m_W &     0        &   0       &  M & \sqrt{2} m_W & 0 \\
0       & - \sqrt{2} m_W    &    M      & 0 & 0 & -\sqrt{2} m_W \\
0 & 0 & \sqrt{2} m_W & 0 & 0 & M \\
0 & 0 & 0 & -\sqrt{2}m_W & M & 0 \\
\end{array}
\right)
\left(
\begin{array}{c}
\chi \\
\tilde{\chi} \\
\omega  \\
\tilde{\omega }\\
\eta \\
\tilde{\eta} \\ 
\end{array}
\right).  \nonumber \\\
\label{MassMatrix}
\eea
The zero modes of the charged fermions have a common Dirac mass of $M$.

To simplify our analysis, we set $m=\tilde{m}$, 
   and in this case we find an approximate expression 
   for the mass eigenvalues of ${\cal M}_{\rm N6}$ by treating $m_W$ terms as perturbation ($m_W \ll |M|$): 
\bea
&&m_1 \simeq  M - m + \frac{2 m_W^2}{2M-m}, \nonumber \\
&&m_2 = m_3 \simeq M + \frac{m_W^2}{M}, \nonumber \\
&&m_4 \simeq M + \frac{m_W^2}{M} + \frac{2M_W^2}{2M+m},  \nonumber \\
&&m_5 \simeq M + \frac{m_W^2}{M} + \frac{2m_W^2}{2M-m}, \nonumber \\
&&m_6 \simeq M + m + \frac{2 m_W^2}{2M+m}, 
\label{MassEigenvalues6}
\eea 
for the mass eigenstates defined as 
 $ U_{{\cal M}6} \; (\chi,  \tilde{\chi}, \omega, \tilde{\omega}, \eta, \tilde{\eta})^T = (\phi_1, \phi_2, \phi_3, \phi_4, \phi_5, \phi_6)^T $  
 with a unitary matrix 
\bea
&&
U_{{\cal M}6}=
\frac{1}{\sqrt{2}}
\left(
\begin{array}{cccccc}
 1 &  0 &  0  & 0 &  a & 0 \\
0 & 1 & 0 & 0  & -c & 0  \\
 0  & 0  & 1 & c & 0 & 0 \\
 0  & 0 & -c & 1 & 0 & -b \\
-a & c & 0 & 0 & 1 & 0 \\
0 & 0 & b & 0 &0 & 1 \\
\end{array}
\right)
\left(
\begin{array}{cccccc}
 1 & -1 &  0  &  0  & 0 & 0 \\
   0  & 0  &  0  & 0 & 1 & -1 \\
  0  &  0  & 1  & -1 & 0 & 0 \\
  0  &  0  &  0  & 0 & 1 & 1 \\
  0 & 0 & 1 & 1 & 0 & 0 \\
  1 & 1 & 0 & 0 & 0 & 0 \\
\end{array}
\right), 
\eea 
where 
\bea 
\label{abc6}  
  a = \frac{\sqrt{2}m_W}{m-2M},  \quad  
  b = \frac{\sqrt{2}m_W}{m+2M},  \quad  
  c = \frac{\sqrt{2}m_W}{2M}.   
  \eea
Note that without loss of generality, we can take $M, m \geq 0$.   
Considering the current experimental constraints from the search for an exotic charged fermion, 
   we may take $M \gtrsim 1~{\rm TeV} \gg m_W$ \cite{PDG}. 
In this case, the lowest mass eigenvalue (dark matter mass $m_{\rm DM}$) is given by $|m_1|$. 
 From the explicit form of the mass matrix ${\cal M}_{\rm N6}$ in Eq.~(\ref{MassMatrix}) and $M \gg m_W$, 
   we notice two typical cases for the constituent of the DM particle: 
   (i) the DM particle is mostly composed of SM singlets when $m=\tilde{m} \lesssim M$, or 
   (ii) the DM particle is mostly composed of a linear combination of the components 
    in the SM $SU(2)$ doublets and triplets when $m=\tilde{m} \gtrsim M$.    
In the case (i), the DM particle communicates with the SM particle essentially through the SM Higgs boson.   
On the other hand, the DM particle is quite similar to the so-called Higgsino/wino-like neutralino DM 
   in the minimal supersymmetric SM for the case (ii).  
Since the Higgsino-like neutralino DM has been very well-studied in many literatures (see, for example, \cite{ArkaniHamed:2006mb}), 
   we focus on the case (i) in this paper. 
Note that the case (i) is a realization of the so-called Higgs-portal DM from the GHU scenario. 
We emphasize that in our scenario, the Yukawa couplings in the original Lagrangian 
   are not free parameters, but are identical to the SM $SU(2)$ gauge coupling,  
   thanks to the structure of the GHU scenario. 
This is in sharp contrast to this class of fermion DM scenarios \cite{FDM}, 
   where the Yukawa couplings are all free parameters.

Now we describe the coupling between the DM particle and the Higgs boson. 
In the original basis, the interaction can be read off from Eq.~(\ref{MassMatrix}) 
   by $v \to v+h$ as 
\bea
{\cal L}_{{\rm Higgs-coupling}} &=&  
-\frac{1}{2} 
\left( \frac{\sqrt{2}m_W}{v} \right) h 
\left(
\begin{array}{cccccc}
\overline{\chi} & \overline{\tilde{\chi}} & \overline{\omega} & \overline{\tilde{\omega}} & \overline{\eta} & \overline{\tilde{\eta}} 
\end{array}
\right) 
{\cal M}_{h6}
\left(
\begin{array}{c}
\chi \\
\tilde{\chi} \\
\omega  \\
\tilde{\omega }\\
\eta \\
\tilde{\eta} \\
\end{array}
\right)  \nonumber \\
&=&
-\frac{1}{2} 
\left( \frac{\sqrt{2}m_W}{v} \right) h 
\left(
\begin{array}{cccccc}
\overline{\phi_1} & \overline{\phi_2} & \overline{\phi_3} & \overline{\phi_4} & \overline{\phi_5} & \overline{\phi_6} \\
\end{array}
\right)
{\cal C}_{h6}
\left(
\begin{array}{c}
\phi_1 \\
\phi_2 \\
\phi_3 \\
\phi_4\\
\phi_5\\
\phi_6\\
\end{array}
\right),  
\label{Hint6}
\eea
where $h$ is the physical Higgs boson, 
and the explicit form of the matrix ${\cal M}_{h6}$ and ${\cal C}_{h6}$ are given by 
\bea
{\cal M}_{h6} 
\equiv
\left(
\begin{array}{cccccc}
0   &   0   &  1  & 0 & 0 & 0 \\
0  &   0    &   0  & -1 & 0 & 0 \\
1  &   0    &   0   &  0 & 1 & 0 \\
0  &  -1   &   0   & 0 & 0 & -1 \\
0  & 0    & 1    & 0 & 0 & 0 \\
0  & 0   &  0   & -1 & 0 & 0\\
\end{array}
\right), 
\eea
\bea
{\cal C}_{h6}  \equiv 
\left(
\begin{array}{cccccc}
-2 a & -a+c & 0 & 0 & -1-ac+a^2 & 0 \\
-a+c & 2c & 0 & 0 & -1-ac+c^2 & 0 \\
0 & 0 & -2 c & -1+b+c^2 & 0 & -1-bc \\
0 & 0 & -1+b+c^2 & 2(1-b)c & 0 & -b+b^2 + c \\
-1-ac+a^2  & -1-ac+c^2 &  0  &  0  & 2(a-c) & 0 \\
0 & 0 & -1-bc & -b+b^2+c & 0 & -2b \\
\end{array}
\right). 
\nonumber \\
\label{Hphys6} 
\eea
The interaction Lagrangian relevant to the DM physics is given by 
\bea
{\cal L}_{{\rm DM-H6}} &=& 
   \frac{\sqrt{2}m_W}{2M-m} \left( \frac{\sqrt{2} m_W}{v} \right)  \; h \; \overline{\psi_{{\rm DM}}} \; \psi_{{\rm DM}}  
  +\left[ 
  \left(\frac{\sqrt{2}m_W}{2M-m} + \frac{m_W}{\sqrt{2}M} \right) \left(\frac{\sqrt{2}m_W}{v} \right)  \; h \overline{\phi_2} \; \psi_{{\rm DM}} 
 \right.  \nonumber \\
 &&  \left. 
 +\left(-1 + \frac{2m_W^2}{(2M-m)^2} + \frac{m_W^2}{(2M-m)M} \right) \left(\frac{\sqrt{2}m_W}{v} \right)  \; h \overline{\phi_5} \; \psi_{{\rm DM}} 
 + {\rm h.c.} \right] ,    
\label{DMHcoupling6}
\eea
where we have identified the lightest mass eigenstate $\phi_1$ as the DM particle ($\psi_{\rm DM}$). 

\subsection{${\bf 10}$-plet MDM}
Next we consider that case that the DM particle is provided by a pair of ${\bf 10}$-plets. 
The Lagrangian relevant to our DM physics discussion is given by
\bea
{\cal L}_{{\rm DM10}} &=& {\rm Tr} \left[ \overline{\psi}({\bf 10}) \; i D\!\!\!\!/ \; \psi({\bf 10}) 
  + \overline{\tilde{\psi}}({\bf 10}) \;  i D\!\!\!\!/  \; {\tilde \psi}({\bf 10}) )
  - M (\overline{\psi}({\bf 10}) \tilde{\psi}({\bf 10}) 
  + \overline{\tilde{\psi}}({\bf 10}) \psi({\bf 10}) ) \right]
  \nonumber \\
  &&+ \delta(y) \left[ 
  \frac{m}{2} \overline{\nu_{sR}^{(0)c}} \nu_{sR}^{(0)} 
  +\frac{\tilde{m}}{2} \overline{\tilde{\nu}_{sL}^{(0)c}} \tilde{\nu}_{sL}^{(0)} 
  + {\rm h.c.} \right], 
\label{DMLagrangian10}
\eea
 where $D\!\!\!\!/$ is the covariant derivative. 
It is useful to express the ${\bf 10}$-plet  fermion in a following matrix form,  
\bea
\psi_{ijk}({\bf 10})= \psi_{1jk} + \psi_{2jk} + \psi_{3jk}, 
\eea
where
\bea
&&
\psi_{1jk} \equiv
\left(
\begin{array}{ccc}
\Delta^{++} & \frac{1}{\sqrt{3}} \Delta^+ & \frac{1}{\sqrt{3}} \Sigma_\uparrow \\
\frac{1}{\sqrt{3}}\Delta^+ & \frac{1}{\sqrt{3}} \Delta^0 & \frac{1}{\sqrt{6}}\Sigma_0 \\
\frac{1}{\sqrt{3}}\Sigma_\uparrow & \frac{1}{\sqrt{6}}\Sigma_0 & \frac{1}{\sqrt{3}}\nu \\
\end{array}
\right), \quad
\psi_{2jk} \equiv
\left(
\begin{array}{ccc}
\frac{1}{\sqrt{3}}\Delta^{+} & \frac{1}{\sqrt{3}} \Delta^0 & \frac{1}{\sqrt{6}}\Sigma_0 \\
\frac{1}{\sqrt{3}}\Delta^0 & \Delta^- & \frac{1}{\sqrt{3}}\Sigma_\downarrow \\
\frac{1}{\sqrt{6}}\Sigma_0 & \frac{1}{\sqrt{3}}\Sigma_\downarrow & \frac{1}{\sqrt{3}}e \\
\end{array}
\right),  \nonumber \\
&&
\psi_{3jk} \equiv
\left(
\begin{array}{ccc}
\frac{1}{\sqrt{3}}\Sigma^\uparrow & \frac{1}{\sqrt{6}}\Sigma_0 & \frac{1}{\sqrt{3}}\nu \\
\frac{1}{\sqrt{6}}\Sigma_0 & \frac{1}{\sqrt{3}}\Sigma_\downarrow & \frac{1}{\sqrt{3}}e \\
\frac{1}{\sqrt{3}}\nu & \frac{1}{\sqrt{3}}e & \nu_s \\
\end{array}
\right). 
\label{rep10-1}
\eea
The quantum numbers for $SU(2)$ representation, 
 the third component of the isospin $I_3$, and the electric charge $Q_{{\rm em}}$ 
 of newly introduced fermions $\Delta$ in the matrix are 
\bea
\Delta^{++}({\bf 4}, 3/2, 3),  \quad
\Delta^+({\bf 4}, 1/2, 2),  \quad
\Delta^0({\bf 4}, -1/2, 1), \quad
\Delta^-({\bf 4}, -3/2, 0).  
\label{rep10-2}
\eea
The corresponding mirror fermion ${\bf 10}$-plet  $\tilde{\psi}({\bf 10})$ takes a similar form 
 with tilde for all fields.

With the non-trivial orbifold boundary conditions, 
   the bulk $SU(3)$ ${\bf 10}$-plet fermions are decomposed into 
   the SM $SU(2)$ quartet, triplet, doublet and singlet fermions. 
The DM particle is provided as a linear combination of 
   the $SU(2)$ singlet, doublet, triplet and quartet components in the ${\bf 10}$-plet  fermions. 
In Eq.~(\ref{DMLagrangian10}), as in the case of ${\bf 6}$-plet fermions, 
   we have also introduced a bulk mass ($M$) to avoid exotic massless fermions  
   and Majorana mass terms on the brane at $y=0$ 
   for the zero modes of the SM singlet components of the ${\bf 10}$-plets ($(\nu_s^{(0)})_R$ and $(\tilde{\nu}_s^{(0)})_L$)    
   to vanish its spin-independent cross section with nucleons 
   through the $Z$-boson exchange in the non-relativistic limit.

Let us focus on the following terms in Eq.~(\ref{DMLagrangian10}), 
   which are relevant to the mass terms in 4-dimensional effective theory: 
\bea
{\cal L}_{{\rm mass10}} &=& 
 {\rm Tr} \left[ \overline{\psi} i \Gamma^5 (\partial_y - 3 ig \langle A_y \rangle) \psi 
 + \overline{\tilde{\psi}} i \Gamma^5 (\partial_y -3 ig \langle A_y \rangle) \tilde{\psi} 
 - M(\overline{\psi} \tilde{\psi} + \overline{\tilde{\psi}} \psi) \right]
 \nonumber \\
 &&+ \delta(y) \left[ \frac{m}{2} \overline{\nu_{sR}^{(0)c}} \nu_{sR}^{(0)} 
 +\frac{\tilde{m}}{2} \overline{\tilde{\nu}_{sL}^{(0)c}} \tilde{\nu}_{sL}^{(0)}  
 + {\rm h.c.} \right]. 
\label{L10_mass}
\eea
Expanding the bulk fermions in terms of KK modes by noticing $Z_2$ parity of $\Delta$ as
\bea
&&\Delta_L(-y) = + \Delta_L(y), \quad  \tilde{\Delta}_L(-y) = -\tilde{\Delta}_L(y), \\
&&\Delta_R(-y) = - \Delta_R(y), \quad  \tilde{\Delta}_R(-y) = +\tilde{\Delta}_R(y) 
\eea
and integrating out the fifth coordinate $y$, we obtain the expression in 4-dimensional effective theory.  
The zero-mode parts for the electric-charge neutral fermions are found to be 
\bea
{\cal L}^{{\rm zero-mode}}_{{\rm mass10}} 
&=& 
 3i m_W \left[ 
 \frac{1}{\sqrt{3}} \overline{\Delta_L^{-(0)}} \Sigma_{\downarrow R}^{(0)} 
 + \overline{e_L} \left( \frac{2}{3} \Sigma_{\downarrow R} + \frac{1}{\sqrt{3}} \nu_{sR} \right) 
+ \frac{1}{\sqrt{3}} \overline{\tilde{\Sigma}_{\downarrow L}} \tilde{\Delta}^-_R 
\nonumber \right. \\
 && \left. 
 + \left( \frac{2}{3} \overline{\tilde{\Sigma}_{\downarrow L}^{(0)}} + \frac{1}{\sqrt{3}} \overline{\tilde{\nu}_{sL}^{(0)}} \right) \tilde{e}_R^{(0)} 
 + {\rm h.c.} \right] 
 \nonumber \\
 && -M \left( \overline{\Delta^-_L} \tilde{\Delta}_R^- 
 + \overline{\Sigma_{\downarrow R}^{(0)}} \tilde{\Sigma}_{\downarrow L}^{(0)} 
 + \overline{\nu_{sR}^{(0)}} \tilde{\nu}_{sL}^{(0)} + \overline{\tilde{e}_R^{(0)}} e_L^{(0)} 
+{\rm h.c.} \right)  
\nonumber \\
&& +
\left( \frac{m}{2} \overline{\nu_{sR}^{(0)c}} \nu_{sR}^{(0)} 
 + \frac{\tilde{m}}{2} \overline{\tilde{\nu}_{sL}^{(0)c}} \tilde{\nu}_{sL}^{(0)} 
 + {\rm h.c.} \right) 
\nonumber \\
& \to & 
 -3 m_W \left[ 
 \frac{1}{\sqrt{3}} \overline{\Delta_L^{-(0)}} \Sigma_{\downarrow R}^{(0)} 
 + \overline{e_L} \left( \frac{2}{3} \Sigma_{\downarrow R} + \frac{1}{\sqrt{3}} \nu_{sR} \right) 
+ \frac{1}{\sqrt{3}} \overline{\tilde{\Sigma}_{\downarrow L}} \tilde{\Delta}^-_R 
\nonumber \right. \\
 && \left. 
 + \left( \frac{2}{3} \overline{\tilde{\Sigma}_{\downarrow L}^{(0)}} + \frac{1}{\sqrt{3}} \overline{\tilde{\nu}_{sL}^{(0)}} \right) \tilde{e}_R^{(0)} 
 + {\rm h.c.} \right] 
 \nonumber \\
 && -M \left( \overline{\Delta^-_L} \tilde{\Delta}_R^- 
 + \overline{\Sigma_{\downarrow R}^{(0)}} \tilde{\Sigma}_{\downarrow L}^{(0)} 
 + \overline{\nu_{sR}^{(0)}} \tilde{\nu}_{sL}^{(0)} + \overline{\tilde{e}_R^{(0)}} e_L^{(0)} 
+{\rm h.c.} \right)  
\nonumber \\
&& -
\left( \frac{m}{2} \overline{\nu_{sR}^{(0)c}} \nu_{sR}^{(0)} 
 + \frac{\tilde{m}}{2} \overline{\tilde{\nu}_{sL}^{(0)c}} \tilde{\nu}_{sL}^{(0)} 
 + {\rm h.c.} \right) , 
\eea
where the arrow means the phase rotations defined in (\ref{phaserot}). 
Rewriting these mass terms in a Majorana basis defined in Rq.~(\ref{Majorana})  and 
\bea
\xi \equiv \Delta_L^{(0)-} + (\Delta_L^{(0)-})^c, \quad 
\tilde{\xi} \equiv \tilde{\Delta}_R^{(0)-} + (\tilde{\Delta}_R^{(0)-})^c, 
\label{Majoranabasis10}
\eea 
we have
\bea
{\cal L}_{10{\rm mass}}^{{\rm zero-mode}} 
 &=& 
-\frac{1}{2}(
\begin{array}{cccccccc}
\overline{\chi} & \overline{\tilde{\chi}} & \overline{\omega} & \overline{\tilde{\omega}} & \overline{\eta} & \overline{\tilde{\eta}} 
& \overline{\xi} & \overline{\tilde{\xi }} \\
\end{array}
)
{\cal M}_{{\rm N}10}
\left(
\begin{array}{c}
\chi \\
\tilde{\chi} \\
\omega  \\
\tilde{\omega }\\
\eta \\
\tilde{\eta} \\
\xi \\
\tilde{\xi} \\
\end{array}
\right) , 
\eea
where
\bea
{\cal M}_{{\rm N}10} \equiv 
\left( 
\begin{array}{cccccccc}
m                   & M                      &  \sqrt{3} m_W & 0                       & 0                    & 0                      & 0                     & 0 \\
M                   & {\tilde m}           &   0                    & - \sqrt{3} m_W & 0                    & 0                      & 0                    & 0 \\
\sqrt{3} m_W &     0                   &   0                   &  M                     & 2 m_W           & 0                      & 0                     & 0 \\
0                    & - \sqrt{3} m_W  &    M                & 0                        & 0                    & -2 m_W            & 0                     & 0 \\
0                    & 0                       & 2 m_W           & 0                        & 0                    & M                      & \sqrt{3} m_W & 0 \\
0                    & 0                       & 0                     & -2 m_W             & M                    & 0                      & 0                    & -\sqrt{3} m_W \\
0                    & 0                       & 0                     & 0                        & \sqrt{3} m_W & 0                      & 0                     & M \\
0                    & 0                       & 0                     & 0                        & 0                     & -\sqrt{3} m_W & M                    & 0 \\
\end{array}
\right). \nonumber \\
\label{MassMatrix10}
\eea
The zero modes of the charged fermions have a common Dirac mass of $M$.

As in the case of ${\bf 6}$-plet, we set $m=\tilde{m}$ to simplify our analysis, 
   and we also find an approximate expression 
   for the mass eigenvalues of ${\cal M}_{{\rm N}10}$  in the case of $m_W \ll |M|$: 
\bea
&&m_1 \simeq  M - m + \frac{3 m_W^2}{2M-m}, \nonumber \\
&&m_2 \simeq M + \frac{1}{2} \left[ \frac{3m_W^2}{2M+m} + \frac{7m_W^2}{M} 
-\sqrt{\left( \frac{3m_W^2}{2M+m} + \frac{m_W^2}{2M} \right)^2 + \frac{12m_W^4}{M^2}} \right], \nonumber \\
&&m_3 \simeq M + \frac{1}{2} \left[ \frac{3m_W^2}{2M-m} + \frac{7m_W^2}{M} 
-\sqrt{\left( \frac{3m_W^2}{2M-m} + \frac{m_W^2}{2M} \right)^2 + \frac{12m_W^4}{M^2}} \right], \nonumber \\
&&m_4 = m_5 \simeq M + \frac{7m_W^2}{2M},  \nonumber \\
&&m_6 \simeq M + \frac{1}{2} \left[ \frac{3m_W^2}{2M+m} + \frac{7m_W^2}{M} 
+\sqrt{\left( \frac{3m_W^2}{2M+m} + \frac{m_W^2}{2M} \right)^2 + \frac{12m_W^4}{M^2}} \right], \nonumber \\
&&m_7 \simeq M + \frac{1}{2} \left[ \frac{3m_W^2}{2M-m} + \frac{7m_W^2}{M} 
+\sqrt{\left( \frac{3m_W^2}{2M-m} + \frac{m_W^2}{2M} \right)^2 + \frac{12m_W^4}{M^2}} \right], \nonumber \\
&&m_8 \simeq M + m + \frac{3 m_W^2}{2M+m}, 
\label{MassEigenvalues10}
\eea 
for the mass eigenstates defined as 
 $ U_{{\cal M}10} \; (\chi,  \tilde{\chi}, \omega, \tilde{\omega}, \eta, \tilde{\eta}, \xi, \tilde{\xi})^T 
 = (\phi_1, \phi_2, \phi_3, \phi_4, \phi_5, \phi_6, \phi_7, \phi_8)^T $  
 with a unitary matrix 
\bea
&&
U_{{\cal M}10}=
\frac{1}{\sqrt{2}}
\left(
\begin{array}{cccccccc}
1   &  0 & 0 &  0 &  0  & 0   & a' & 0 \\
0   &  1 & 0 &  0 & 0   & -d' & 0  & 0 \\
0   &  0 & 1 &  0 & -d' & 0  & -c' & 0 \\
0   &  0 & 0 &  1 & 0  & -c' & 0  & -b' \\
0   &  0 & d' & 0 & 1  & 0   & 0  & 0 \\
0   & d' & 0 &  c' & 0 & 1  & 0  & 0 \\
-a' &  0 & c' & 0 & 0  & 0  & 1  & 0 \\
0   &  0 & 0 &  b' & 0 & 0  & 0  & 1 \\
\end{array}
\right)
\left(
\begin{array}{cccccccc}
 1  & -1 &  0  &  0  & 0 &  0 & 0 & 0 \\
 0  & 0  &  0  &  0  & 0 &  0 & 1 & -1 \\
 0  & 0  &  0  &  0  & 1 & -1 & 0 & 0 \\
 0  & 0  &  1  & -1  & 0 & 0 & 0 & 0 \\
 0  & 0  &  0 &  0  & 0 & 0 & 1 & 1 \\
 0  & 0  &  0 &  0  & 1 & 1  & 0 & 0 \\
 0  & 0  &  1 &  1  & 0 & 0 & 0 & 0 \\
 1 &  1 &  0  &  0 & 0 &  0 & 0 & 0 \\
\end{array}
\right), \nonumber \\
\eea 
where 
\bea 
\label{abc10}  
  a' = \frac{\sqrt{3}m_W}{m-2M},  \quad  
  b' = \frac{\sqrt{3}m_W}{m+2M},  \quad  
  c' = \frac{m_W}{M}, \quad 
  d'= \frac{\sqrt{3}m_W}{2M}. 
\eea
As in the case of ${\bf 6}$-plets,  
   we  take $M \gtrsim 1~{\rm TeV} \gg m_W$ \cite{PDG}, and the DM mass is given by $m_{\rm DM}=|m_1|$. 
For $m=\tilde{m} \gtrsim M$,  the DM particle is mostly composed of a linear combination of 
   the components in the $SU(2)$ doublets, triplets and quartets. 
Since this case is also very similar to the Higgsino/wino-like neutralino DM, 
   we focus on the Higgs-portal DM case with $m=\tilde{m} \lesssim M$ 
   also for this ${\bf 10}$-plet case.

We describe the coupling between the DM particle and the Higgs boson in the ${\bf 10}$-plet case. 
In the original basis, the interaction can be read off from Eq.~(\ref{MassMatrix10}) 
   by $v \to v+h$ as 
\bea
{\cal L}_{{\rm Higgs-coupling}} &=&  
-\frac{1}{2} 
\left( \frac{m_W}{v} \right) h 
\left(
\begin{array}{cccccccc}
\overline{\chi} & \overline{\tilde{\chi}} & \overline{\omega} & \overline{\tilde{\omega}} & \overline{\eta} & \overline{\tilde{\eta}} & \overline{\xi} & \overline{\tilde{\xi}} 
\end{array}
\right) 
{\cal M}_{h10}
\left(
\begin{array}{c}
\chi \\
\tilde{\chi} \\
\omega  \\
\tilde{\omega }\\
\eta \\
\tilde{\eta} \\
\xi \\
\tilde{\xi} \\
\end{array}
\right)  \nonumber \\
&=&
-\frac{1}{2} 
\left( \frac{m_W}{v} \right) h 
\left(
\begin{array}{cccccccc}
\overline{\phi_1} & \overline{\phi_2} & \overline{\phi_3} & \overline{\phi_4} & \overline{\phi_5} & \overline{\phi_6} & \overline{\phi_7} & \overline{\phi_8} \\
\end{array}
\right)
{\cal C}_{h10}
\left(
\begin{array}{c}
\phi_1 \\
\phi_2 \\
\phi_3 \\
\phi_4\\
\phi_5\\
\phi_6\\
\phi_7\\
\phi_8\\
\end{array}
\right),  
\label{Hint10}
\eea
where $h$ is the physical Higgs boson, 
and the explicit form of the matrix ${\cal M}_{h10}$ and ${\cal C}_{h10}$ are given by 
\bea
{\cal M}_{h10} 
\equiv
\left(
\begin{array}{cccccccc}
0   &   0   &  \sqrt{3}  & 0 & 0 & 0 & 0 & 0 \\
0  &   0    &   0  & -\sqrt{3} & 0 & 0 & 0 & 0 \\
\sqrt{3}  &   0    &   0   &  0 & 2 & 0 & 0 & 0 \\
0  &  -\sqrt{3}   &   0   & 0 & 0 & -2 & 0 & 0 \\
0  & 0    & 2    & 0 & 0 & 0 & \sqrt{3} & 0 \\
0  & 0   &  0   & -2 & 0 & 0 & 0 & -\sqrt{3} \\
0 & 0 & 0 & 0 & \sqrt{3} & 0 & 0 & 0 \\
0 & 0 & 0 & 0 & 0 & -\sqrt{3} & 0 & 0 \\
\end{array}
\right), 
\eea
and 
\bea
{\cal C}_{h10}  &\equiv& 
\left(
\begin{array}{cccccccc}
{\cal C}_{11}  & 0                      &  {\cal C}_{13}  & 0                     & {\cal C}_{15} & 0                     & {\cal C}_{17} & 0 \\
0                      & {\cal C}_{22}  & 0                       & {\cal C}_{24} & 0                     & {\cal C}_{26} & 0                     & {\cal C}_{28} \\
{\cal C}_{31}  & 0                      & {\cal C}_{33}   & 0                     & {\cal C}_{35} & 0                     & {\cal C}_{37} & 0 \\
0                      & {\cal C}_{42} & 0                        & {\cal C}_{44} & 0                     & {\cal C}_{46} & 0                     & {\cal C}_{48} \\
{\cal C}_{51}  & 0                     & {\cal C}_{53}    & 0                     & {\cal C}_{55} &  0                    & {\cal C}_{57} & 0 \\
0                      & {\cal C}_{62} & 0                        & {\cal C}_{64} & 0                     & {\cal C}_{66} & 0                     & {\cal C}_{68} \\
{\cal C}_{71}  & 0                     & {\cal C}_{73}    & 0                     & {\cal C}_{75} & 0                     & {\cal C}_{77} & 0 \\
0                     & {\cal C}_{82}  & 0                        & {\cal C}_{84} & 0                     & {\cal C}_{86} & 0                     & {\cal C}_{88} \\
\end{array}
\right),
\eea
with  
\bea
&&
{\cal C}_{11} = -2\sqrt{3}a', \quad  
{\cal C}_{22} = 2\sqrt{3}d', \quad   
{\cal C}_{33} = 2(2 c' + \sqrt{3}d'), \quad   
{\cal C}_{44} = 2(\sqrt{3}b' + 2c'), \nonumber \\
&&
{\cal C}_{55} = -2\sqrt{3}d', \quad  
{\cal C}_{66} = -2(2c' + \sqrt{3}d'), \quad 
{\cal C}_{77} = 2(\sqrt{3}a' -2c'), \quad 
{\cal C}_{88} = -2\sqrt{3}b', \nonumber \\ 
&&
{\cal C}_{13} = {\cal C}_{31} = -2a' + \sqrt{3}c', \quad  
{\cal C}_{15} = {\cal C}_{51} = -2a'd',\quad  
{\cal C}_{17} = {\cal C}_{71} = -2a'c' + \sqrt{3}(-1+a'^2), \quad 
\nonumber \\
&&
{\cal C}_{24} = {\cal C}_{42} = \sqrt{3}c' + 2d', \quad  
{\cal C}_{26} = {\cal C}_{62} =  2c'd' + \sqrt{3}(-1+d'^2), \quad 
{\cal C}_{28} = {\cal C}_{82} = 2b'd', \quad 
\nonumber \\
&&
{\cal C}_{35} = {\cal C}_{53} = 2c'd' + \sqrt{3}(-1 + d'^2), \quad  
{\cal C}_{37} = {\cal C}_{73} = 2(1-c'^2) + \sqrt{3}c'(-a'+d'), \quad 
\nonumber \\
&&
{\cal C}_{46} = {\cal C}_{64} =  2(-1+c'^2) + \sqrt{3}c'(b'+d'), \quad 
{\cal C}_{48} = {\cal C}_{84} = \sqrt{3}(-1+b'^2) + 2b'c', \quad 
\nonumber \\
&&
{\cal C}_{57} = {\cal C}_{75} = -\sqrt{3}c' -2d', \quad   
{\cal C}_{68} = {\cal C}_{86} = -2b' -\sqrt{3}c'. 
\label{Hphys} 
\eea
The interaction Lagrangian relevant to the DM physics can be read as 
\bea
{\cal L}_{{\rm DM-H10}} &=& 
  \frac{2\sqrt{3}m_W}{2M-m} \left( \frac{\sqrt{3}m_W}{v} \right)  \; h \; \overline{\psi_{{\rm DM}}} \; \psi_{{\rm DM}}  \nonumber \\
 && + \left[
   \left(\frac{2m_W}{2M-m} + \frac{m_W}{M} \right) \left(\frac{\sqrt{3}m_W}{v} \right)  \; h \overline{\phi_3} \; \psi_{{\rm DM}} 
 +\frac{\sqrt{3}m_W^2}{(2M-m)M} \left(\frac{\sqrt{3}m_W}{v} \right)  \; h \overline{\phi_5} \; \psi_{{\rm DM}} 
 \right. \nonumber \\
 && \left. 
 + \left( -1 + \frac{3m_W^2}{(2M-m)^2} + \frac{2m_W^2}{(2M-m)M} \right) \left(\frac{\sqrt{3}m_W}{v} \right)  \; h \overline{\phi_7} \; \psi_{{\rm DM}} 
   + {\rm h.c.} \right] ,    
\label{DMHcoupling10}
\eea
where we have identified the lightest mass eigenstate $\phi_1$ as the DM particle ($\psi_{\rm DM}$).

\subsection{${\bf 15}$-plet MDM}
It is straightforward to extend our previous analysis further 
  higher-representations. 
In this subsection, we finally consider a pair of ${\bf 15}$-plet bulk fermions, ${\psi}({\bf 15})$ and  $\tilde{\psi}({\bf 15})$.  
The Lagrangian relevant to our DM physics discussion is given by
\bea
{\cal L}_{{\rm DM15}} &=& {\rm Tr} \left[ \overline{\psi}({\bf 15}) \; i D\!\!\!\!/ \; \psi({\bf 15}) 
  + \overline{\tilde{\psi}}({\bf 15}) \;  i D\!\!\!\!/  \; {\tilde \psi}({\bf 15}) )
  - M (\overline{\psi}({\bf 15}) \tilde{\psi}({\bf 15}) 
  + \overline{\tilde{\psi}}({\bf 15}) \psi({\bf 15}) ) \right]
  \nonumber \\
  &&+ \delta(y) \left[ 
  \frac{m}{2} \overline{\nu_{sR}^{(0)c}} \nu_{sR}^{(0)} 
  +\frac{\tilde{m}}{2} \overline{\tilde{\nu}_{sL}^{(0)c}} \tilde{\nu}_{sL}^{(0)} 
  + {\rm h.c.} \right].  
\label{DMLagrangian15}
\eea
It is useful to express the ${\bf 15}$-plet in a following matrix form,  
\bea
\psi_{ijkl}({\bf 15})= \psi_{11kl} + \psi_{12kl} + \psi_{13kl} + \psi_{22kl} + \psi_{23kl} + \psi_{33kl}, 
\eea
where
\bea
&&
\psi_{11kl} \equiv
\left(
\begin{array}{ccc}
\Theta_{\uparrow \uparrow} & \frac{1}{2} \Theta_\uparrow & \frac{1}{2\sqrt{3}} \Delta^{++} \\
\frac{1}{2}\Theta_\uparrow & \frac{1}{\sqrt{6}} \Theta_0 & \frac{1}{2\sqrt{3}}\Delta^+ \\
\frac{1}{2}\Delta^{++} & \frac{1}{2\sqrt{3}}\Delta^+ & \frac{1}{\sqrt{6}} \Sigma_\uparrow \\
\end{array}
\right), \quad
\psi_{12kl} \equiv
\left(
\begin{array}{ccc}
\frac{1}{2}\Theta_{\uparrow} & \frac{1}{\sqrt{6}} \Theta_0 & \frac{1}{2\sqrt{3}} \Delta^+ \\
\frac{1}{\sqrt{6}}\Theta_0 & \frac{1}{2} \Theta_\downarrow & \frac{1}{2\sqrt{3}}\Delta^0 \\
\frac{1}{2\sqrt{3}}\Delta^+ & \frac{1}{2\sqrt{3}}\Delta^0 & \frac{1}{2\sqrt{3}} \Sigma_0 \\
\end{array}
\right),  \nonumber \\
&&
\psi_{13kl} \equiv
\left(
\begin{array}{ccc}
\frac{1}{2}\Delta^{++} & \frac{1}{2\sqrt{3}}\Delta^+ & \frac{1}{\sqrt{6}} \Sigma_\uparrow \\
\frac{1}{2\sqrt{3}}\Delta^+ & \frac{1}{2\sqrt{3}}\Delta_0 & \frac{1}{2\sqrt{3}}\Sigma_0 \\
\frac{1}{\sqrt{6}}\Sigma_\uparrow & \frac{1}{2\sqrt{3}}\Sigma_0 & \frac{1}{2}\nu \\
\end{array}
\right), \quad
\psi_{22kl} \equiv
\left(
\begin{array}{ccc}
\frac{1}{\sqrt{6}} \Theta_0 & \frac{1}{2} \Theta_\downarrow & \frac{1}{2\sqrt{3}} \Delta^0 \\
\frac{1}{2}\Theta_\uparrow & \Theta_{\downarrow \downarrow} & \frac{1}{2} \Delta^- \\
\frac{1}{2\sqrt{3}}\Delta^{0} & \frac{1}{2}\Delta^- & \frac{1}{\sqrt{6}} \Sigma_\downarrow \\
\end{array}
\right), 
\nonumber \\
&&
\psi_{23kl} \equiv
\left(
\begin{array}{ccc}
\frac{1}{2\sqrt{3}}\Delta^{+} & \frac{1}{2\sqrt{3}} \Delta^0 & \frac{1}{2\sqrt{3}} \Sigma_0 \\
\frac{1}{\sqrt{3}}\Delta^0 & \frac{1}{2} \Delta^- & \frac{1}{\sqrt{6}}\Sigma_{\downarrow} \\
\frac{1}{2\sqrt{3}}\Sigma_0 & \frac{1}{\sqrt{6}}\Sigma_{\downarrow} & \frac{1}{2} e \\
\end{array}
\right),  \quad
\psi_{33kl} \equiv
\left(
\begin{array}{ccc}
\frac{1}{\sqrt{6}}\Sigma_{\uparrow} & \frac{1}{2\sqrt{3}}\Sigma_0 & \frac{1}{2} \nu \\
\frac{1}{2\sqrt{3}}\Sigma_0 & \frac{1}{\sqrt{6}}\Sigma_\downarrow & \frac{1}{2}e \\
\frac{1}{2}\nu & \frac{1}{2}e & \nu_s \\
\end{array}
\right).
\eea
The quantum numbers for $SU(2)$ representation, 
 the third component of the isospin $I_3$, and the electric charge $Q_{{\rm em}}$ 
 of newly introduced fermions $\Theta$ in the matrix are 
\bea
\Theta_{\uparrow \uparrow}({\bf 5}, 2, 4),  \quad
\Theta_\uparrow({\bf 5}, 1, 3),  \quad
\Theta_0({\bf 5}, 0, 1), \quad
\Theta_\downarrow({\bf 5}, -1, 1), \quad   
\Theta_{\downarrow \downarrow}({\bf 5}, -2, 0). 
\eea
The corresponding mirror fermion ${\bf 15}$-plet $\tilde{\psi}({\bf 15})$ takes a similar form 
 with tilde for all fields.

With the non-trivial orbifold boundary conditions, 
   the bulk $SU(3)$ ${\bf 15}$-plet fermions are decomposed into 
   the SM $SU(2)$ quintet, quartet, triplet, doublet and singlet fermions. 
Let us focus on the mass terms in Eq.~(\ref{DMLagrangian15}): 
\bea
{\cal L}_{{\rm mass15}} &=& 
 {\rm Tr} \left[ \overline{\psi} i \Gamma^5 (\partial_y - 4 ig \langle A_y \rangle) \psi 
 + \overline{\tilde{\psi}} i \Gamma^5 (\partial_y -4 ig \langle A_y \rangle) \tilde{\psi} 
 - M(\overline{\psi} \tilde{\psi} + \overline{\tilde{\psi}} \psi) \right]
 \nonumber \\
 &&+ \delta(y) \left[ \frac{m}{2} \overline{\nu_{sR}^{(0)c}} \nu_{sR}^{(0)} 
 +\frac{\tilde{m}}{2} \overline{\tilde{\nu}_{sL}^{(0)c}} \tilde{\nu}_{sL}^{(0)}  
 + {\rm h.c.} \right]. 
\label{L15_mass}
\eea
Expanding the bulk fermions in terms of KK modes by noticing $Z_2$ parity of $\Delta$ as
\bea
&&\Theta_L(-y) = - \Theta_L(y), \quad  \tilde{\Theta}_L(-y) = + \tilde{\Theta}_L(y), \\
&&\Theta_R(-y) = + \Theta_R(y), \quad  \tilde{\Theta}_R(-y) = - \tilde{\Theta}_R(y) 
\eea
and integrating out the fifth coordinate $y$, we obtain the expression in 4-dimensional effective theory.  
The zero-mode parts for the electric-charge neutral fermions are found to be 
\bea
{\cal L}^{{\rm zero-mode}}_{{\rm mass15}} 
&=& 
 -4i m_W \left[ 
 \frac{1}{2} \overline{\Delta^{-}_L} \Theta_{\downarrow \downarrow R} 
 + \frac{3}{2\sqrt{6}} \overline{\Delta^{-}_L} \Sigma_{\downarrow R}  
 + \frac{3}{2\sqrt{6}} \overline{e_L} \Sigma_{\downarrow R} 
 + \frac{1}{2} \overline{e_L} \nu_{sR} 
 + \frac{1}{2} \overline{\tilde{\Delta}^-_R} \tilde{\Theta}_{\downarrow \downarrow L} 
 \right. \nonumber \\
 && \left. 
 + \frac{3}{2\sqrt{6}} \overline{\tilde{\Delta}^-_{R}} \tilde{\Sigma}_{\downarrow \downarrow L} 
 + \frac{3}{2\sqrt{6}} \overline{\tilde{e}_R} \tilde{\Sigma}_{\downarrow L} 
 + \frac{1}{2} \overline{\tilde{e}_R} \tilde{\nu_{sL}} 
 + {\rm h.c.} \right] 
 \nonumber \\
 && -M \left( 
 \overline{\Theta_{\downarrow \downarrow R}} \tilde{\Theta}_{\downarrow \downarrow L}
 + \overline{\Delta^-_L} \tilde{\Delta}_R^- 
 + \overline{\Sigma_{\downarrow R}} \tilde{\Sigma}_{\downarrow L} 
 + \overline{e_L} \tilde{e_R} 
 + \overline{\nu_{sR}} \tilde{\nu}_{sL} 
+{\rm h.c.} \right)  
\nonumber \\
&& +
\left( \frac{m}{2} \overline{\nu_{sR}^{(0)c}} \nu_{sR}^{(0)} 
 + \frac{\tilde{m}}{2} \overline{\tilde{\nu}_{sL}^{(0)c}} \tilde{\nu}_{sL}^{(0)} 
 + {\rm h.c.} \right) 
\nonumber \\
& \to & 
 -4 m_W \left[ 
 \frac{1}{2} \overline{\Delta^{-}_L} \Theta_{\downarrow \downarrow R} 
 + \frac{3}{2\sqrt{6}} \overline{\Delta^{-}_L} \Sigma_{\downarrow R}  
 + \frac{3}{2\sqrt{6}} \overline{e_L} \Sigma_{\downarrow R} 
 + \frac{1}{2} \overline{e_L} \nu_{sR} 
 + \frac{1}{2} \overline{\tilde{\Delta}^-_R} \tilde{\Theta}_{\downarrow \downarrow L} 
\right. \nonumber \\
&& \left. 
 + \frac{3}{2\sqrt{6}} \overline{\tilde{\Delta}^-_{R}} \tilde{\Sigma}_{\downarrow \downarrow L} 
 + \frac{3}{2\sqrt{6}} \overline{\tilde{e}_R} \tilde{\Sigma}_{\downarrow L} 
 + \frac{1}{2} \overline{\tilde{e}_R} \tilde{\nu_{sL}} 
 + {\rm h.c.} \right] 
 \nonumber \\
 && -M \left( 
 \overline{\Theta_{\downarrow \downarrow R}} \tilde{\Theta}_{\downarrow \downarrow L}
 + \overline{\Delta^-_L} \tilde{\Delta}_R^- 
 + \overline{\Sigma_{\downarrow R}} \tilde{\Sigma}_{\downarrow L} 
 + \overline{e_L} \tilde{e_R} 
 + \overline{\nu_{sR}} \tilde{\nu}_{sL} 
+{\rm h.c.} \right)  
\nonumber \\
&& -
\left( \frac{m}{2} \overline{\nu_{sR}^{(0)c}} \nu_{sR}^{(0)} 
 + \frac{\tilde{m}}{2} \overline{\tilde{\nu}_{sL}^{(0)c}} \tilde{\nu}_{sL}^{(0)} 
 + {\rm h.c.} \right) , 
\eea
where the arrow means the phase rotations defined in (\ref{phaserot}) and 
\bea
\Theta_{\downarrow \downarrow} \to i \Theta_{\downarrow \downarrow}, \quad 
\tilde{\Theta}_{\downarrow \downarrow} \to i \tilde{\Theta}_{\downarrow \downarrow}.  
\eea 
Rewriting these mass terms in a Majorana basis with Eq.~(\ref{Majorana}),  Eq.~(\ref{Majoranabasis10}) and   
\bea
\rho \equiv (\Theta^{(0)}_{\downarrow \downarrow})_R + (\Theta_{\downarrow \downarrow}^{(0)})_R^c, 
\quad
\tilde{\rho} \equiv (\tilde{\Theta}^{(0)}_{\downarrow \downarrow})_L + (\tilde{\Theta}_{\downarrow \downarrow}^{(0)})_L^c,
\eea
we have the following mass terms, 
\bea
{\cal L}_{15{\rm mass}}^{{\rm zero-mode}} 
 &=& 
-\frac{1}{2}(
\begin{array}{cccccccccc}
\overline{\chi} & \overline{\tilde{\chi}} & \overline{\omega} & \overline{\tilde{\omega}} & \overline{\eta} & \overline{\tilde{\eta}} 
& \overline{\xi} & \overline{\tilde{\xi }} & \overline{\rho} & \overline{\tilde{\rho}} \\
\end{array}
)
{\cal M}_{{\rm N}15}
\left(
\begin{array}{c}
\chi \\
\tilde{\chi} \\
\omega  \\
\tilde{\omega }\\
\eta \\
\tilde{\eta} \\
\xi \\
\tilde{\xi} \\
\rho \\
\tilde{\rho} \\
\end{array}
\right)  , 
\eea
where
\bea
&&{\cal M}_{{\rm N}15} \equiv 
\nonumber \\
&&
\left( 
\begin{array}{cccccccccc}
m             & M                 &   2m_W          & 0                       & 0                      & 0                       & 0                     & 0                         & 0         & 0 \\
M            & {\tilde m}     &   0                     & -2m_W            & 0                      & 0                       & 0                    & 0                          & 0         & 0 \\
2m_W    &     0              &   0                    &  M                     & \sqrt{6} m_W  & 0                       & 0                     & 0                         & 0         & 0 \\
0             & -2 m_W      &    M                   & 0                        & 0                     & -\sqrt{6} m_W  & 0                     & 0                        & 0          & 0 \\
0             & 0                  & \sqrt{6} m_W  & 0                        & 0                     & M                       & \sqrt{6} m_W & 0                       & 0           & 0 \\
0             & 0                  & 0                      & -\sqrt{6} m_W   & M                    & 0                       & 0                      & -\sqrt{6} m_W & 0           & 0 \\
0             & 0                  & 0                      & 0                        & \sqrt{6} m_W & 0                       & 0                      & M                      & 2m_W  & 0 \\
0             & 0                  & 0                      & 0                        & 0                     & -\sqrt{6} m_W  & M                     & 0                       & 0           & -2m_W \\
0             &0                   &0                       &0                          &0                     &0                         &2m_W             &0                         &0            &M \\
0             &0                   &0                       &0                         &0                      &0                         &0                       &-2m_W              & M           &0 \\
\end{array}
\right). \nonumber \\
\label{MassMatrix15}
\eea
The zero modes of the charged fermions have a common Dirac mass of $M$.

\section{Dark Matter Relic Abundance}
\label{RA}
In this section, we evaluate the DM relic abundance and identify an allowed parameter region 
  to reproduce the observed DM relic density \cite{Planck2015} (68 \% confidence level):  
\bea
\Omega_{{\rm DM}} h^2 = 0.1198 \pm 0.0015. 
\eea
In our model, the DM physics is controlled by only two free parameters, namely, $m$ and $M$. 
As we discussed in the previous section, we focus on the Higgs-portal DM case 
  with $0 \leq m \lesssim M$. 
In this section, we consider the ${\bf 6}$-plet and ${\bf 10}$-plet cases. 
Analysis for a more higher-dimensional representation is analogous to those presented in this section. 

According to the interaction Lagrangian in Eqs.~(\ref{DMHcoupling6}) and (\ref{DMHcoupling10}), 
  we consider two main annihilation processes of a pair of DM particles as analyzed in \cite{MMOO}. 
One is through the $s$-channel Higgs boson exchange, and 
  the other is the process $\psi_{\rm DM} \psi_{\rm DM} \to h h$ through 
  the exchange of $\phi_5$ $(\phi_7)$ in the $t/u$-channel for the ${\bf 6}$-plet (${\bf 10}$-plet) case, 
  when the DM particle is heavier than the SM Higgs boson.  
In evaluating this process, we may use an effective Lagrangian of the form, 
\bea 
 {\cal L}_{\rm DM-H}^{\rm eff} = 
 \left\{
 \begin{array}{c}
 \frac{1}{2} \left(\frac{\sqrt{2} m_W}{v}\right)^2 \frac{1}{m_5} \;
    h \; h \; \overline{\psi_{\rm DM}} \; \psi_{\rm DM}~({\bf 6}{\textrm -}{\rm plet}), \\
     \frac{1}{2} \left(\frac{\sqrt{3} m_W}{v}\right)^2 \frac{1}{m_7} \;
    h \; h \; \overline{\psi_{\rm DM}} \; \psi_{\rm DM}~({\bf 10}{\textrm -}{\rm plet}) 
 \end{array}
\right.
\eea 
which is obtained by integrating $\phi_5$ $(\phi_7)$ out, 
  and calculate the DM pair annihilation cross section times relative velocity ($v_{\rm rel}$) as 
\bea 
 \sigma v_{\rm rel} = 
 \left\{
 \begin{array}{l}
 \frac{1}{64 \pi} \left(\frac{\sqrt{2} m_W}{v}\right)^4 \left(\frac{1}{m_5}\right)^2 v_{\rm rel}^2
 \equiv \sigma_0 v_{\rm rel}^2~({\bf 6}{\textrm -}{\rm plet}), \\
 \frac{1}{64 \pi} \left(\frac{\sqrt{3} m_W}{v}\right)^4 \left(\frac{1}{m_7}\right)^2 v_{\rm rel}^2
 \equiv \sigma_0 v_{\rm rel}^2~({\bf 10}{\textrm -}{\rm plet}). 
\end{array}
\right.
\eea
It is well-known that the observed DM relic density is reproduced by $\sigma_0 \sim 1$ pb.  
Since we find $\sigma_0 \sim 0.04$ $(0.06)$ pb for 
 $m_5$ $(m_7) \simeq M=1$ TeV in the case of ${\bf 6}$-plet (${\bf 10}$-plet), 
  we conclude that the observed relic density is not reproduced by the process $\psi_{\rm DM} \psi_{\rm DM} \to h h$.

Next we consider the DM pair annihilation through the $s$-channel Higgs boson exchange 
   when the DM particle is lighter than the Higgs boson.  
Since the coupling between the pair of DM particles and the Higgs boson is suppressed,  
   an enhancement of the DM annihilation cross section through the Higgs boson resonance 
   is necessary to reproduce the observed relic DM density. 
We evaluate the DM relic abundance by integrating the Boltzmann equation 
\bea
\frac{dY}{dx} = - \frac{xs\langle \sigma v\rangle}{H(m_{\rm DM})} (Y^2-Y^2_{\rm EQ}),
\label{Boltzmanneq}
\eea
where the temperature of the universe is normalized by the DM mass as $x=m_{\rm DM}/T$, 
 $H(m_{{\rm DM}})$ is the Hubble parameter as $T=m_{\rm DM}$, 
 $Y$ is the yield (the ratio of the DM number density to the entropy density $s$) of the DM particle, 
 $Y_{{\rm EQ}}$ is the yield of the DM in thermal equilibrium, 
 and $\langle \sigma v_{\rm rel} \rangle$ is the thermal average of 
   the DM annihilation cross section times relative velocity for a pair of the DM particles.  
Various quantities in the Boltzmann equation are given as follows. 
\bea
s=\frac{2\pi^2}{45}g_* \frac{m^3_{{\rm DM}}}{x^3}, 
\quad 
H(m_{{\rm DM}}) =\sqrt{\frac{\pi^2}{90} g_*} \frac{m^2_{{\rm DM}}}{M_P}, 
\quad  
sY_{{\rm EQ}} = \frac{g_{{\rm DM}}}{2\pi^2} \frac{m^3_{{\rm DM}}}{x} K_2(x),
\eea 
where $M_P = 2.44 \times 10^{18}$ GeV is the reduced Planck mass, 
 $g_{{\rm DM}}=2$ is the number of degrees of freedom for the DM particle, 
 $g_*$ is the effective total number of degrees of freedom for the particles in thermal equilibrium 
  (in our analysis, we use $g_{*}=86.25$ corresponding to $m_{\rm DM}\simeq m_h/2$ 
  with the Higgs boson mass of 125 GeV), 
  and $K_2$ is the modified Bessel function of the second kind. 
For $m_{\rm DM} \simeq m_h/2 = 62.5$ GeV, a DM pair annihilates into a pair of the SM fermions as  
   $\psi_{{\rm DM}} \psi_{{\rm DM}} \to h \to f \bar{f}$, where $f$ denotes the SM fermions. 
We calculate the cross section for the annihilation process as 
\bea
\sigma(s) =\frac{y^2_{\rm DM}}{16\pi} 
\left[ 3 \left( \frac{m_b}{v} \right)^2 + 3 \left( \frac{m_c}{v} \right)^2 + \left( \frac{m_\tau}{v} \right)^2 \right] 
\frac{\sqrt{s(s - 4m^2_{\rm DM})}}{(s-m_h^2)^2 + m_h^2 \Gamma_h^2}, 
\eea
where 
$y_{\rm DM}$ can be lead from Eqs.~(\ref{DMHcoupling6}) and (\ref{DMHcoupling10}) as
\bea
y_{{\rm DM}} \simeq
\left\{
\begin{array}{l}
\frac{2\sqrt{2} m_W}{2M-m} \frac{\sqrt{2}m_W}{v}~({\bf 6}{\textrm-}{\rm plet}) \\
\frac{4\sqrt{3} m_W}{2M-m} \frac{\sqrt{3}m_W}{v}~({\bf 10}{\textrm -}{\rm plet}), \\
\end{array}
\right.
\eea  
  and we have only considered pairs of bottom, charm and tau for the final states, 
  neglecting the other lighter quarks, and used the following values for the fermion masses 
  at the $Z$-boson mass scale \cite{Bora:2012tx}:  
  $m_b=2.82$ GeV, $m_c=685$ MeV and $m_\tau=1.75$ GeV. 
The total Higgs boson decay width $\Gamma_h$ is given by 
  $\Gamma_h = \Gamma_h^{\rm SM} + \Gamma_h^{\rm new}$, 
  where $\Gamma_h^{\rm SM}=4.07$ MeV \cite{SMGamma_h} 
  is the total Higgs boson decay width in the SM and 
\bea
\Gamma_h^{{\rm new}} =
\left\{
\begin{array}{cc}
0 & m_h < 2 m_{\rm DM} \\
\frac{m_h}{16\pi} \left( 1- \frac{4m_{{\rm DM}}^2}{m_h^2} \right)^{3/2} y_{{\rm DM}}^2 & m_h > 2 m_{\rm DM} \\
\end{array}
\right.,
\eea
  is the partial decay width of the Higgs boson to a DM pair. 
The thermal average of the annihilation cross section is given by
\bea
\langle \sigma v \rangle = (sY_{{\rm EQ}})^{-2} g_{{\rm DM}}^2 
\frac{m_{{\rm DM}}}{64 \pi^4 x} \int_{4m_{{\rm DM}}}^\infty ds 
\hat{\sigma}(s) \sqrt{s} K_1 \left(\frac{x \sqrt{s}}{m_{{\rm DM}}} \right), 
\eea
where $\hat{\sigma}(s) = 2(s-4m_{\rm DM}^2) \sigma(s)$ is the reduced cross section 
 with the total annihilation cross section $\sigma(s)$, 
 and $K_1$ is the modified Bessel function of the first kind. 
We solve the Boltzmann equation numerically 
  and find an asymptotic value of the yield $Y(\infty)$ to obtain the present DM relic density as
\bea
\Omega h^2  = \frac{m_{{\rm DM}}s_0 Y(\infty)}{\rho_c/h^2},
\eea
where $s_0=2890$ cm$^{-3}$ is the entropy density of the present universe,  
 and $\rho_c/h^2=1.05 \times 10^{-5}$ GeV/cm$^3$ is the critical density.

\begin{figure}[t]
\centering
\begin{center}
\includegraphics[width=0.7\textwidth,angle=0,scale=1.05]{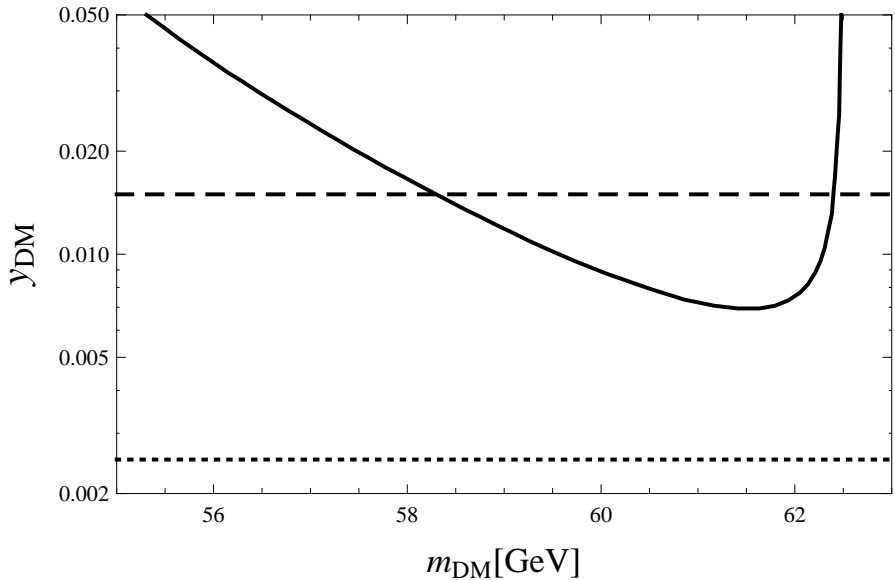}
\end{center}
  \caption{
The Yukawa coupling $y_{\rm DM}$ as a function of $m_{\rm DM}$ (solid line) 
   along which the observed DM relic density $\Omega_{\rm DM} h^2=0.1198$ is reproduced.  
Here, the current experimental upper bound from the XENON 1T result \cite{XENON1T} and 
   the prospective reach in the future LUX-ZEPLIN DM experiment \cite{LZ}
   are also shown as the dashed and the dotted lines, respectively.  
}
\label{CB}
\end{figure}

In Fig.~\ref{CB}, we show $y_{\rm DM}$ as a function of $m_{\rm DM}$ (solid line) 
   along which the observed DM relic density $\Omega_{\rm DM} h^2=0.1198$ is reproduced.  
Here, the current experimental upper bound from the XENON 1T result \cite{XENON1T} 
  and the prospective reach in the future LUX-ZEPLIN DM experiment \cite{LZ} 
  are also shown as the dashed and the dotted lines, respectively, 
  which will be derived in Sec.~\ref{DirectD}.  
In order to satisfy the XENON 1T constraint, 
   we find the parameter regions such as 
   $58.3 \leq m_{\rm DM}[{\rm GeV}] \leq 62.4$ and $(0.00692 \leq)\; y_{\rm DM} \leq 0.0150$. 
Using Eqs.~(\ref{MassEigenvalues6}) and (\ref{MassEigenvalues10}), 
  we can express $M$ as a function of $m_{\rm DM}$ along the solid line in Fig.~\ref{CB}. 
Our results are shown in the Fig.~\ref{MvsmDM}. 
Corresponding to the parameter regions of 
  $58.3 \leq m_{\rm DM}[{\rm GeV}] \leq 62.4$ and $(0.00692 \leq)\; y_{\rm DM} \leq 0.0150$,  
  we have found $6.97 \leq M[{\rm TeV}] \; (\leq 15.1)$ and    
  $10.5 \leq M[{\rm TeV}] \; (\leq 22.6)$ 
  for the ${\bf 6}$-plet and ${\bf 10}$-plet cases, respectively. 

\begin{figure}[t]
\begin{center}
\includegraphics[width=0.465\textwidth,angle=0,scale=1.05]{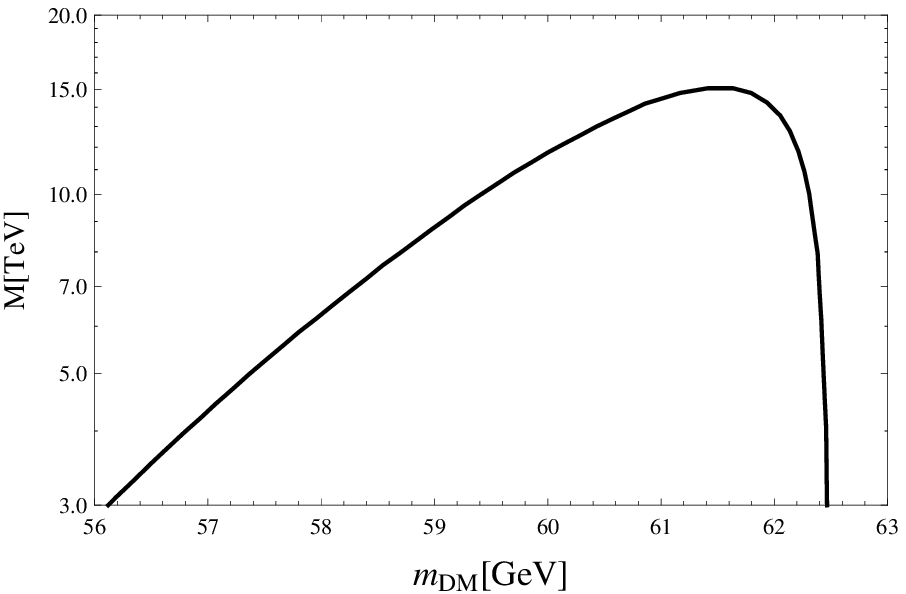}
\hspace{0.1cm}
\includegraphics[width=0.46\textwidth,angle=0,scale=1.05]{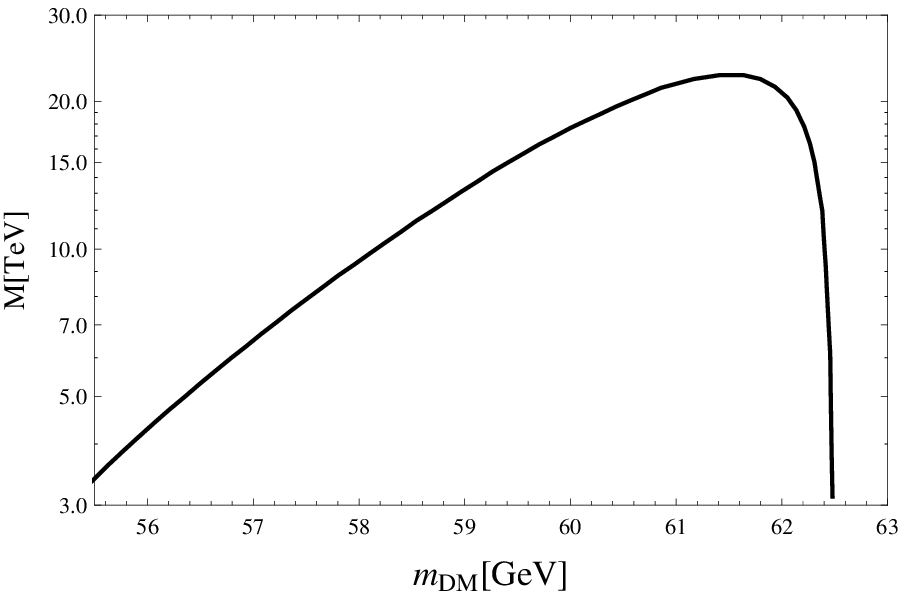}
\end{center}
\caption{
$M$ as a function of $m_{\rm DM}$
   for the ${\bf 6}$-plet case (left panel) and the ${\bf10}$-plet case (right panel),  
   along the solid line in Fig.~\ref{CB}. 
}
\label{MvsmDM}
\end{figure}

\section{Direct Dark Matter Detection Constraints} 
\label{DirectD}

Currently, many experiments are in operation and also planned 
  for directly detecting a dark matter particle through its elastic scattering off with nuclei. 
In this section, we calculate the spin-independent elastic scattering cross section 
  of the DM particle via the Higgs boson exchange 
  and derive the constraint on the model parameters 
  from the current experimental results.

The spin-independent elastic scattering cross section with nucleon is given by 
\bea
\sigma_{{\rm SI}} = 
\frac{1}{\pi}
\left(
\frac{y_{{\rm DM}}}{v}
\right)^2
\left(
\frac{\mu_{\psi_{{\rm DM}}N}}{m_h^2}
\right)^2
f_N^2, 
\label{DD}
\eea
where $\mu_{\psi_{{\rm DM}}N} = m_N m_{{\rm DM}}/(m_N+m_{{\rm DM}})$ is 
   the reduced mass of the DM-nucleon system with the nucleon mass $m_N=0.939$ GeV, and 
\bea
f_N = 
\left(
\sum_{q=u,d,s} f_{T_q} + \frac{2}{9}f_{TG}
\right) 
m_N
\eea
   is the nuclear matrix element accounting for the quark and gluon contents of the nucleon. 
In evaluating $f_{T_q}$, we employ the results from the lattice QCD simulation \cite{LatQCD} 
(see also \cite{QCD2}):  
   $f_{T_u} +f_{T_d} \simeq 0.056$ and $|f_{T_s}|\leq 0.08$. 
To make our analysis conservative, we set $f_{T_s}=0$.  
Using the trace anomaly formula, $\sum_{q=u,d,s} f_{T_q} + f_{TG}=1$ \cite{TraceAnomaly}, 
   we obtain $f_N^2 \simeq 0.0706 \; m_N^2$, and hence the spin-independent elastic scattering cross section 
   is approximately given by 
\bea
 \sigma_{{\rm SI}} \simeq 
4.47 \times 10^{-7} \; {\rm pb} \times y_{\rm DM}^2 
\label{DD2}
\eea
 for $m_{\rm DM}=m_h/2=62.5$ GeV.

For the upper bound on the spin-independent cross section,
   we refer the recent XENON 1T result \cite{XENON1T}: $\sigma_{\rm SI} \leq 1.0 \times 10^{-10}$ pb 
   for $m_{\rm DM} \simeq 62.5$ GeV. 
 From Eq.~(\ref{DD2}), we find $y_{\rm DM} \leq 0.0150$, 
   which is depicted as the horizontal dashed line in Fig.~\ref{CB}. 
The next-generation successor of the LUX experiment, the LUX-ZEPLIN experiment \cite{LZ}, 
   plans to improve the current upper bound on the spin-independent cross section 
   by about two orders of magnitude.  
When we apply the search reach of the LUX-ZEPLIN experiment 
   as $\sigma_{\rm SI} \leq 2.8 \times 10^{-12}$ pb,
   we obtain $y_{\rm DM} \leq 0.00251$. 
This prospective upper bound is shown as the dotted line in Fig.~\ref{CB}. 
The present allowed parameter region will be all covered by the future LUX-ZEPLIN experiment.

\section{125 GeV Higgs boson mass}
\label{Hmass}
In this section, we reproduce the 125 GeV Higgs boson mass in our GHU model 
   in the presence of the bulk fermion multiplets. 
For our calculations, we follow a 4-dimensional effective theory approach of 
  the 5-dimensional GHU scenario developed in Ref.~\cite{GHcondition}.  
According to this approach, a low-energy effective Higgs quartic coupling in the GHU model 
  can be easily obtained by solving the renormalization group (RG) equation 
  of the Higgs quartic coupling with a vanishing quartic coupling at the compactification scale. 
This so-called ``gauge-Higgs condition" \cite{GHcondition} corresponds to 
  a restoration of the 5-dimensional gauge invariance at the compactification scale.  
For energies higher than the compactification scale, 
  our theory behaves as the original 5-dimensional theory and no Higgs potential exists there.

We have two mass parameters important in our RG analysis, namely, 
   the bulk mass $M \simeq m$ and the compactification scale $M_{\rm KK}=1/R$. 
Although the electroweak symmetry breaking causes the mass splittings among the bulk fermion zero modes, 
    the mass splittings are all negligibly small due to  $M \simeq m \gg m_W$. 
Thus, all mass eigenstates except for two mass eigenstates 
    ($\phi_1$ and $\phi_6$ ($\phi_1$ and $\phi_8$) in the ${\bf 6}$-plet (${\bf 10}$-plet) case)  
   are degenerate. 
The non-degenerate mass eigenstates are mostly composed of the SM singlets, and 
    we safely neglect their contributions to our RG analysis.

For the renormalization scale smaller than the bulk mass $\mu < M$, 
   all bulk fermions are decoupled and we employ the SM RG equations at the two-loop level~\cite{RGE} 
   (See \cite{MMOO} for explicit the SM RG equations.).
In solving these RGEs, we use the boundary conditions at the top quark pole mass ($M_t$)
  given in \cite{RGE_Higgs_quartic}: 
\bea 
g_1(M_t)&=&\sqrt{\frac{5}{3}} \left( 0.35761 + 0.00011 (M_t - 173.10) 
   -0.00021  \left( \frac{M_W-80.384}{0.014} \right) \right),  
\nonumber \\
g_2(M_t)&=& 0.64822 + 0.00004 (M_t - 173.10) + 0.00011  \left( \frac{M_W-80.384}{0.014} \right) , 
\nonumber \\
g_3(M_t)&=& 1.1666 + 0.00314 \left(  \frac{\alpha_s-0.1184}{0.0007}   \right) ,
\nonumber \\
y_t(M_t) &=& 0.93558 + 0.0055 (M_t - 173.10) - 0.00042  \left(  \frac{\alpha_s-0.1184}{0.0007}   \right) 
\nonumber \\
  && - 0.00042 \left( \frac{M_W-80.384}{0.014} \right) ,
\nonumber \\
\lambda(M_t) &=&  2 (0.12711 + 0.00206 (m_h - 125.66) - 0.00004 (M_t - 173.10) ) . 
\eea 
We employ $M_W=80.384$ GeV, $\alpha_s =0.1184$, 
   the central value of the combination of Tevatron and LHC measurements of top quark mass 
   $M_t=173.34$ GeV~\cite{top_pole_mass}, and the central value of the updated Higgs boson mass measurement, 
   $m_h=125.09$ GeV from the combined analysis by the ATLAS  and the CMS collaborations~\cite{Higgs_Mass_LHC}.


For the renormalization scale $\mu \geq M$, 
   the SM RG equations are modified in the presence of the bulk fermions.  
In this paper, we take only one-loop corrections from the bulk fermions into account.  
For the case with a pair of the bulk {\bf 6}-plet fermions, 
   the beta functions of the $SU(2)$ and $U(1)_Y$ gauge couplings receive 
   new contributions as 
\bea
\Delta b_1= 2 \left(\frac{2}{3} + \frac{24}{5} Q^2 \right), \; \; \; 
\Delta b_2 =\frac{20}{3}, 
\eea
where $Q=2/3$ is the $U(1)^\prime$ charge of the bulk {\bf 6}-plet fermions. 
The beta functions of the top Yukawa and Higgs quartic couplings are modified as 
\bea 
&& \beta_t^{(1)} \to \beta_t^{(1)}  + 2 y_t  \left(2 |Y_S|^2 + 3 |Y_D|^2 \right),  \nonumber \\
&& \beta_{\lambda}^{(1)} \to \beta_{\lambda}^{(1)} +
       2 \left[ \lambda \left( 8 |Y_S|^2 + 12 |Y_D|^2 \right) 
       - \left( 8 |Y_S|^4 + 10 |Y_D|^4 +16 |Y_S|^2 |Y_D|^2 \right) \right].  
\label{LamBeta}
\eea
Here, the Yukawa couplings, $Y_S$ and $Y_D$, are defined in the Yukawa interactions 
 from Eq.~(\ref{DMLagrangian6}) as
 \bea
 {\cal L}_{{\rm DM6}} \supset -Y_S \bar{D}HS -Y_D \bar{D} T H^\dagger,
 \eea
where $S, D$ and $T$ stand for the $SU(2)$ singlet, doublet and triplet fields 
  in the decomposition of ${\bf 6} = {\bf 1} \oplus {\bf 2} \oplus {\bf 3}$ 
  (see Eqs.~(\ref{rep6-1}) and (\ref{rep6-2})).  
The Yukawa couplings obey the following RG equations: 
\bea 
16 \pi^2 \frac{d Y_S}{d \ln \mu} &=&
  Y_S \left[ 3 y_t^2 - \left( \frac{9}{20}  g_1^2  + \frac{9}{4} g_2^2 \right) 
  + 2 \left( \frac{7}{2} |Y_S|^2 + \frac{19}{4}  |Y_D|^2 \right) \right. \nonumber \\
  && -\left.  \frac{18}{5} \left( \frac{2}{3}- Q\right) \left(\frac{1}{6}-Q \right) g_1^2 \right], \nonumber \\
16 \pi^2 \frac{d Y_D}{d \ln \mu}  &=&
   Y_D  \left[ 3 y_t^2  - \left( \frac{9}{20} g_1^2 + \frac{9}{4} g_2^2 \right)
   + 2 \left( \frac{9}{2}  |Y_S|^2 + \frac{17}{4} |Y_D|^2  \right)  \right. \nonumber \\
   &&- \left. 6 g_2^2 - \frac{18}{5} \left(\frac{1}{6} - Q\right) \left(\frac{1}{3} -Q \right) g_1^2  \right]. 
\eea

In our RG analysis, we numerically solve the SM RG equations from $M_t$ to $M$, 
  at which the solutions connect with the solutions of the RG equations with the bulk ${\bf 6}$-plet fermions. 
For a fixed $M$ values, we arrange input values of $|Y_S(M)|$ and $|Y_D(M)|$ so as to find 
  numerical solutions which satisfy the the gauge-Higgs condition and the unification condition 
  among the gauge and Yukawa couplings such that 
\bea 
  \lambda(M_{\rm KK})=0, \; \; \; Y_S(M_{\rm KK})= Y_D(M_{\rm KK}) = - i g_2 (M_{\rm KK}) . 
\eea


For the case with a pair of the bulk ${\bf 10}$-plet fermions, 
  the beta functions of 
  the $SU(2)$ and $U(1)_Y$ gauge couplings receive new contributions as 
\bea
\Delta b_1= 2  \left( 3+ 12 Q^2 \right), \; \; \; 
\Delta b_2 = 20,
\eea
where $Q=1$ is the $U(1)^\prime$ charge of the bulk {\bf 10}-plet fermions. 
The beta functions of the top Yukawa and Higgs quartic couplings are modified as 
\bea 
 \beta_t^{(1)} &\to& \beta_t^{(1)}  + 2 y_t \left(2 |Y_S|^2 + 3 |Y_D|^2 \right),  \nonumber \\
 \beta_{\lambda}^{(1)} &\to& \beta_{\lambda}^{(1)} +
       2 \left[ 4 \lambda \left( 2 |Y_S|^2 + 3 |Y_D|^2 +4 |Y_T|^2\right)  \right. \nonumber \\
& & \left. - \left( 8 |Y_S|^4 + 10 |Y_D|^4 + \frac{112}{9} |Y_T|^4 
                 +16 |Y_S|^2 |Y_D|^2 + \frac{64}{3} |Y_D|^2 |Y_T|^2  \right) \right]. 
\label{LamBeta-10}
\eea
Here, the Yukawa couplings, $Y_S$, $Y_D$ and $Y_T$, are defined in the Yukawa interactions 
 from Eq.~(\ref{DMLagrangian10}) as
 \bea
 {\cal L}_{{\rm DM10}} \supset -Y_S \bar{D}HS -Y_D \bar{D} T H^\dagger -Y_T \bar{F} T H,
 \eea
where $S$, $D$, $T$ and $F$ stand for the $SU(2)$ singlet, doublet, triplet and quartet fields 
  in the decomposition of ${\bf 10} = {\bf 1} \oplus {\bf 2} \oplus {\bf 3} \oplus {\bf 4}$, respectively  
  (see Eqs.~(\ref{rep6-1}), (\ref{rep6-2}), (\ref{rep10-1}) and (\ref{rep10-2})).    
The Yukawa couplings obey the following RG equations: 
\bea 
16 \pi^2 \frac{d Y_S}{d \ln \mu} &=&
  Y_S \left[ 3 y_t^2 - \left( \frac{9}{20}  g_1^2  + \frac{9}{4} g_2^2 \right) 
  + 2 \left( \frac{7}{2} |Y_S|^2 + \frac{27}{4}  |Y_D|^2  + 4 |Y_T|^2  \right) \right. \nonumber \\
  && 
  -\left.  \frac{18}{5} \left(1- Q\right) \left(\frac{1}{2}-Q \right) g_1^2 \right], \nonumber \\
16 \pi^2 \frac{d Y_D}{d \ln \mu}  &=&
   Y_D  \left[ 3 y_t^2  - \left( \frac{9}{20} g_1^2 + \frac{9}{4} g_2^2 \right)
   + 2 \left( \frac{9}{2}  |Y_S|^2 + \frac{17}{4} |Y_D|^2 + \frac{22}{3} |Y_T|^2 \right)  \right. \nonumber \\
   && 
    -\left. 6 g_2^2 - \frac{18}{5} \left(Q-\frac{1}{2} \right) Q \;  g_1^2  \right],  \nonumber \\ 
16 \pi^2 \frac{d Y_T}{d \ln \mu}  &=&
   Y_T  \left[ 3 y_t^2  - \left( \frac{9}{20} g_1^2 + \frac{9}{4} g_2^2 \right)
   + 2 \left( 2  |Y_S|^2 + \frac{11}{2} |Y_D|^2 + \frac{31}{6} |Y_T|^2 \right)  \right. \nonumber \\
   && 
   - \left. 15 g_2^2 - \frac{18}{5} Q \left(Q+\frac{1}{2} \right) g_1^2  \right]. 
\eea

As in the analysis for the bulk ${\bf 6}$-plet fermions, 
  we numerically solve the SM RG equations from $M_t$ to $M$, 
  at which the solutions connect with the solutions of the RG equations with the bulk ${\bf 10}$-plet fermions. 
For a fixed $M$ values, we arrange input values of $|Y_S(M)|$, $|Y_D(M)|$ and $Y_T(M)$ 
  so as to find numerical solutions which satisfy the the gauge-Higgs condition and the unification condition 
  among the gauge and Yukawa couplings such that 
\bea 
  \lambda(M_{\rm KK})=0, \; \; \; 
 \sqrt{\frac{2}{3}} Y_S(M_{\rm KK}) = \sqrt{\frac{1}{2}} Y_D(M_{\rm KK})=   \sqrt{\frac{2}{3}} Y_T(M_{\rm KK})= - i g_2 (M_{\rm KK}) . 
\eea

The RG evolution of Higgs quartic coupling to reproduce the Higgs boson mass 
  $m_H=125.09$ GeV is shown in Fig.~\ref{RGE-Lambda}.    
The dotted line denotes the running quartic coupling in the SM, 
  while the dashed (solid) line corresponds to the result
  for the case with a pair of ${\bf 6}$-plet (${\bf 10}$-plet) bulk fermions. 
For the dashed (solid) line, we find $M=5$ TeV for $M_{\rm KK}=91$ ($8.2$) TeV, 
   at which the gauge-Higgs condition is satisfied. 
When we trace the dashed and solid lines from $M_t$ to higher energies 
   we see that the running of the Higgs quartic coupling is drastically altered 
   from the SM one (dotted line) due to the contributions from the bulk fermions 
   with $M=5$ TeV.

For a fixed $M$ value, we numerically find a $M_{{\rm KK}}$ value. 
The relation between $M$ and $M_{\rm KK}$ is depicted in Fig.~\ref{MkkvsM}. 
In the left (right) panel we show the relation in the ${\bf 6}$-plet (${\bf 10}$-plet) case. 
Thanks to the contributions to the RG equations from the bulk fermions in higher-dimensional representations, 
    no hierarchy between $M$ and $M_{\rm KK}$ is needed to reproduce the 125 GeV Higgs boson mass.  
This is in sharp contrast to the case with the bulk $SU(3)$ triplet fermions considered in Ref.~\cite{MMOO}, 
   where we have found $M_{{\rm KK}}={\cal O}(10^8)$ GeV for $M={\cal O}(1)$ TeV. 
Such a compactification scale is too high for the GHU model to be natural. 
This unnaturalness of the model is significantly relaxed in the presence of the bulk fermions 
   in higher-dimensional representations. 
This is a main point of this paper. 
In Sec.~\ref{RA}, we have found the parameter regions of 
  $6.97 \leq M[{\rm TeV}] \; \leq 15.1$ and $10.5 \leq M[{\rm TeV}] \; \leq 22.6$ 
  for the ${\bf 6}$-plet and ${\bf 10}$-plet cases, respectively, 
  from the constraints on the DM relic abundance and the direct DM detection cross section.  
These $M$ regions correspond to 
  $137  \leq M_{\rm KK}[{\rm TeV}] \; \leq 238$ and 
  $20.2 \leq M_{\rm KK}[{\rm TeV}] \; \leq 41.1$ 
  for the ${\bf 6}$-plet and ${\bf 10}$-plet cases, respectively,

\begin{figure}[htbp]
\centering
   \includegraphics[width=120mm]{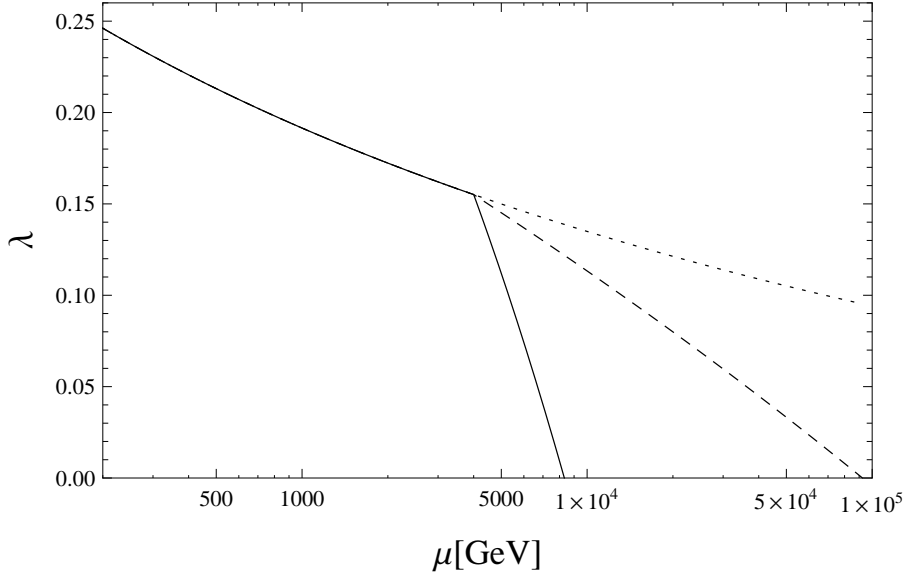}
  \caption{
RG evolution of the Higgs quartic coupling with the bulk mass $M=5$ TeV 
   for the ${\bf 6}$-plet case (dashed line) and the ${\bf 10}$-plet case (solid line), 
   along with the result in the SM (dotted line). 
The compactification scale is found to be 
   $M_{\rm KK}=91$ TeV and $8.2$ TeV for the ${\bf 6}$-plets and  the ${\bf 10}$-plets,  
   respectively, 
   where the gauge-Higgs condition $\lambda(M_{\rm KK})=0, |Y_S(M_{\rm KK})|  = g_2 (M_{\rm KK})/\sqrt{2}$ 
   for the ${\bf 6}$-plets, and 
      $\lambda(M_{\rm KK})=0$,
      $\sqrt{\frac{2}{3}} Y_S(M_{\rm KK}) = \sqrt{\frac{1}{2}} Y_D(M_{\rm KK})=   \sqrt{\frac{2}{3}} Y_T(M_{\rm KK})= - i g_2 (M_{\rm KK})$
    for the ${\bf 10}$-plets are satisfied. 
}
\label{RGE-Lambda}
\end{figure}

\begin{figure}[htbp]
\begin{center}
\includegraphics[width=0.46\textwidth,angle=0,scale=1.05]{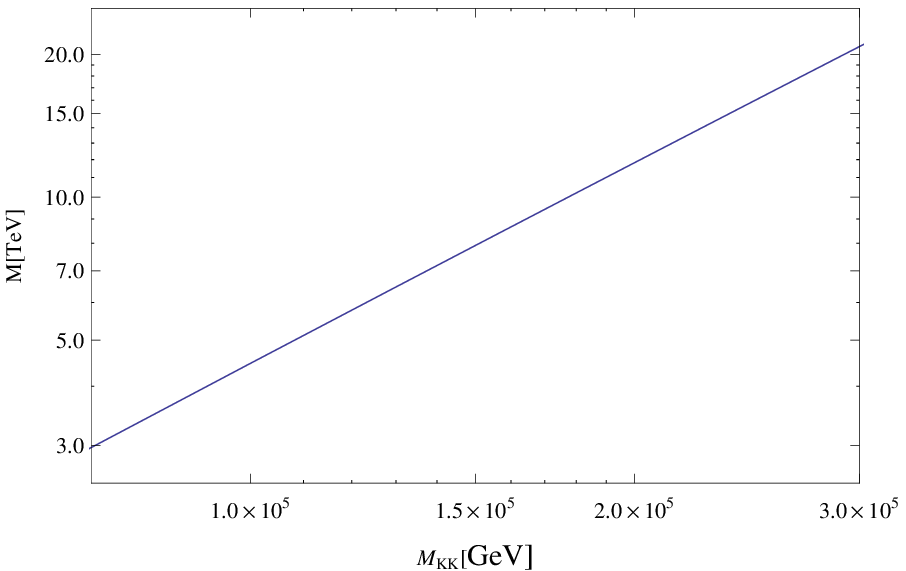}
\hspace{0.1cm}
\includegraphics[width=0.46\textwidth,angle=0,scale=1.05]{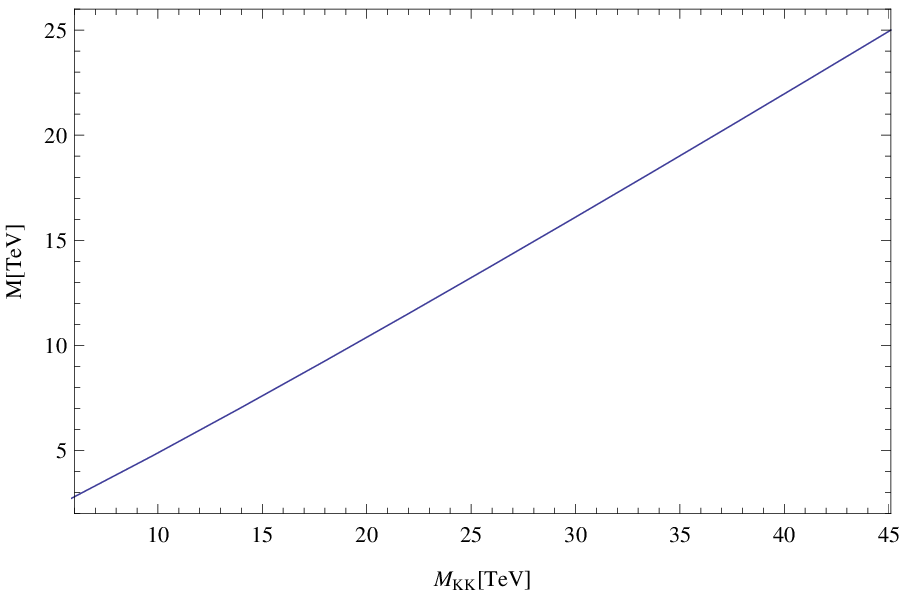}
\end{center}
\caption{
The relation between the bulk mass ($M$) and the compactification scale ($M_{\rm KK}$)
 in the ${\bf 6}$-plet case (left) and ${\bf 10}$-plet case (right)
 so as to reproduce the 125 GeV Higgs boson mass. 
}
\label{MkkvsM}
\end{figure}

\section{Conclusions}
In this paper, we have proposed a MDM scenario 
   in the context of a 5-dimensional GHU model
   based on the gauge group $SU(3) \times U(1)^\prime$ 
   with a compactification of the 5th dimension on $S^1/Z_2$ orbifold. 
We have introduced a pair of bulk $SU(3)$ multiplet fermions in higher-dimensional representations, 
  along with a bulk mass term, Majorana mass terms at an orbifold fixed point, and a periodic boundary condition.  
Associated with the breaking of the $SU(3) \times U(1)^\prime$  into the SM $SU(2) \times U(1)_Y$ gauge group  
  by a non-trivial orbifold boundary condition, 
  the DM particle is provided as a linear combinations of the electric charge neutral components in the multiplets. 
Since the DM particle communicates with the SM particles only through the bulk gauge interactions 
  in the original framework, our model is the GHU version of the MDM scenario. 

We have two typical cases for the constituent of the DM particles. 
One is that the DM particle is mostly a linear combinations of the electric charge neutral components
in the SM $SU(2)$ multiplets. 
The other one is that the DM particle is mostly a singlet under the SM gauge group. 
The first one is very analogous to the Higgsino/wino-like neutralino DM in the minimal supersymmetric SM.  
Since the Higgsino/wino-like neutralino DM has been well studied, 
   we have focused on the latter case, in which the DM particle communicates 
   with the SM particles through the Higgs boson, namely, the Higgs-portal DM scenario. 
As an example, we have investigated the cases with the ${\bf 6}$-plet  and the ${\bf 10}$-plet bulk fermions separately. 
We have identified an allowed parameter region to be consistent with the current experimental constraints, 
   which will be fully covered by the direct dark matter detection experiments in the near future. 
   
Employing the effective theoretical approach with the gauge-Higgs condition, 
  we have also studied the RG evolution of Higgs quartic coupling 
  and shown that the observed Higgs mass of 125 GeV is reproduced  
  with the compactification scale of  $M_{\rm KK}={\cal O}(100)$ TeV (${\cal O}(10)$ TeV) 
  for the case with the ${\bf 6}$-plets (${\bf 10}$-plets), 
  while satisfying the constraints from the DM relic abundance and the XENON 1T result. 
Comparing this result with $M_{\rm KK} \simeq 10^8$ GeV that we have previously obtained 
  in the case with a pari of bulk $SU(3)$ triplet fermions \cite{MMOO}, 
  the unnaturalness of the model with  $M_{\rm KK} \gg 1$ TeV is drastically relaxed 
  thanks to the bulk multiplets in higher-dimensional representations.

Finally, we comment on other possibilities for the realization of DM particle in our model. 
In this paper, we assign the $U(1)^\prime$ charges of $2/3$ and $-1$ 
  for the bulk ${\bf 6}$-plets and ${\bf 10}$-plets, respectively. 
In the SM gauge group decomposition, this charge assignment leads to 
  fermions singlet under the SM gauge group, 
  which are constituents of the DM particle. 
In general, there are other charge assignments to provide electric-charge neutral particle 
   in the SM gauge group decomposition as shown in Table~\ref{MDM}. 
Here, the DM particle is embedded in the SM $SU(2)$ multiplet(s), 
   and this is a realization of the MDM scenario \cite{MDM} 
   as 4-dimensional effective theory of the GHU model.

%
\begin{table}
\begin{center}
\bea
\begin{array}{|c|c|}
\hline
U(1)^\prime & {\bf 6} = {\bf 1} \oplus {\bf 2} \oplus {\bf 3} \\
\hline
\frac{2}{3} & (0)_0 +
\left(
\begin{array}{c}
1 \\
0 \\
\end{array}
\right)_{1/2} +
\left(
\begin{array}{c}
2 \\
1 \\
0 \\
\end{array}
\right)_{1} \\
-\frac{1}{3} & (-1)_{-1} +
\left(
\begin{array}{c}
0 \\
-1 \\
\end{array}
\right)_{-1/2} +
\left(
\begin{array}{c}
1 \\
0 \\
-1 \\
\end{array}
\right)_{0} \\
-\frac{4}{3} & (-2)_{-2} +
\left(
\begin{array}{c}
-1 \\
-2 \\
\end{array}
\right)_{-\frac{3}{2}} +
\left(
\begin{array}{c}
0 \\
-1 \\
-2 \\
\end{array}
\right)_{-1} \\
\hline
& {\bf 10} = {\bf 1} \oplus {\bf 2} \oplus {\bf 3} \oplus {\bf 4} \\
\hline
1 & (0)_0 +
\left(
\begin{array}{c}
1 \\
0 \\
\end{array}
\right)_{\frac{1}{2}} +
\left(
\begin{array}{c}
2 \\
1 \\
0 \\
\end{array}
\right)_{1} +
\left(
\begin{array}{c}
3 \\
2 \\
1 \\
0 \\
\end{array}
\right)_{\frac{3}{2}} \\
0 & 
(-1)_{-1} +
\left(
\begin{array}{c}
0 \\
-1 \\
\end{array}
\right)_{-\frac{1}{2}} +
\left(
\begin{array}{c}
1 \\
0 \\
-1 \\
\end{array}
\right)_{0} +
\left(
\begin{array}{c}
2 \\
1 \\
0 \\
-1 \\
\end{array}
\right)_{\frac{1}{2}} \\
-2 & 
(-2)_{-2} +
\left(
\begin{array}{c}
-1 \\
-2 \\
\end{array}
\right)_{-\frac{3}{2}} +
\left(
\begin{array}{c}
0 \\
-1 \\
-2 \\
\end{array}
\right)_{-1} +
\left(
\begin{array}{c}
1 \\
0 \\
-1 \\
-2 \\
\end{array}
\right)_{-\frac{1}{2}} \\
-3 & 
(-3)_{-3} +
\left(
\begin{array}{c}
-2 \\
-3 \\
\end{array}
\right)_{-\frac{5}{2}} +
\left(
\begin{array}{c}
-1 \\
-2 \\
-3 \\
\end{array}
\right)_{-2} +
\left(
\begin{array}{c}
0 \\
-1 \\
-2 \\
-3 \\
\end{array}
\right)_{-\frac{3}{2}} \\
\hline
\end{array} 
\eea
\end{center}
\caption{
General $U(1)'$ charge assignments providing 
  the DM candidates (electric charge neutral components)
  in the SM $SU(2) \times U(1)_Y$ decomposition of the ${\bf 6}$-plet and ${\bf 10}$-plets 
  of the bulk $SU(3)$ gauge group. 
The numbers in the parenthesis denote the electric charges 
 for the corresponding components of the multiplets. 
Subscripts of the multiplets are their hyper-charges. 
}
 \label{MDM}
\end{table}

\section*{Acknowledgments}
S.O. would like to thank the Department of Physics and Astronomy 
  at the University of Alabama for hospitality during her visit. 
This work is supported in part by JSPS KAKENHI Grant Number JP17K05420 (N.M.) 
  and the DOE Grant Number DE-SC0012447 (N.O.).

\newpage

\end{document}